\newif\ifplainstyle
\plainstyletrue
\plainstylefalse

\newif\ifjhepstyle
\jhepstyletrue

\newif\ifprstyle
\prstyletrue
\prstylefalse

\ifprstyle
	\documentclass[twocolumn,nofootinbib]{revtex4-1}
\else
	\documentclass[11pt,a4paper]{article}
\fi

\ifjhepstyle
	\usepackage{jheppub}
	\usepackage{amsfonts}
	\usepackage{verbatim}
	\usepackage{float}
	\usepackage{color}
	\renewcommand*\arraystretch{2}
	\usepackage{array}
	\newcolumntype{C}[1]{>{\centering\arraybackslash$}p{#1}<{$}}
	\makeatletter
	\def\@fpheader{\phantom{:-)}}
	\makeatother
\else	
 	\ifprstyle
		\usepackage{verbatim}
		\usepackage{amsmath,amsfonts,amssymb}
		\usepackage[colorlinks=true
                	,urlcolor=blue
                	,anchorcolor=blue
                	,citecolor=blue
                	,filecolor=blue
                	,linkcolor=blue
                	,menucolor=blue
                	]{hyperref}
	\else
            	\usepackage{verbatim}
            	\usepackage{cite}
            	\usepackage{setspace}
            	\usepackage[top=2.5cm, bottom=2.75cm, left=2.5cm, right=2.5cm]{geometry}
            	\usepackage{amsmath,amsfonts,amssymb}
            	\usepackage[colorlinks=true
            	,urlcolor=blue
            	,anchorcolor=blue
            	,citecolor=blue
            	,filecolor=blue
            	,linkcolor=blue
            	,menucolor=blue
            	,linktoc=page
            	]{hyperref}
            	\usepackage{float}
            	\restylefloat{table}
            	\renewcommand{\arraystretch}{1.5}
            	\numberwithin{equation}{section}
	\fi
\fi

\allowdisplaybreaks

\usepackage[normalem]{ulem}
\usepackage{tikz}
\usepackage{mathtools}
\usepackage{pgfplots}
\usepackage{subcaption}
\usepackage{bbold}
\usepackage{transparent}
\usepackage{bm}
\usepackage{enumitem}
\usepackage[textsize=scriptsize,textwidth=2.5cm]{todonotes}

\makeatletter
\let\save@mathaccent\mathaccent
\newcommand*\if@single[3]{%
  \setbox0\hbox{${\mathaccent"0362{#1}}^H$}%
  \setbox2\hbox{${\mathaccent"0362{\kern0pt#1}}^H$}%
  \ifdim\ht0=\ht2 #3\else #2\fi
  }
\newcommand*\rel@kern[1]{\kern#1\dimexpr\macc@kerna}
\newcommand*\widebar[1]{\@ifnextchar^{{\wide@bar{#1}{0}}}{\wide@bar{#1}{1}}}
\newcommand*\wide@bar[2]{\if@single{#1}{\wide@bar@{#1}{#2}{1}}{\wide@bar@{#1}{#2}{2}}}
\newcommand*\wide@bar@[3]{%
  \begingroup
  \def\mathaccent##1##2{%
    \let\mathaccent\save@mathaccent
    \if#32 \let\macc@nucleus\first@char \fi
    \setbox\z@\hbox{$\macc@style{\macc@nucleus}_{}$}%
    \setbox\tw@\hbox{$\macc@style{\macc@nucleus}{}_{}$}%
    \dimen@\wd\tw@
    \advance\dimen@-\wd\z@
    \divide\dimen@ 3
    \@tempdima\wd\tw@
    \advance\@tempdima-\scriptspace
    \divide\@tempdima 10
    \advance\dimen@-\@tempdima
    \ifdim\dimen@>\z@ \dimen@0pt\fi
    \rel@kern{0.6}\kern-\dimen@
    \if#31
      \overline{\rel@kern{-0.6}\kern\dimen@\macc@nucleus\rel@kern{0.4}\kern\dimen@}%
      \advance\dimen@0.4\dimexpr\macc@kerna
      \let\final@kern#2%
      \ifdim\dimen@<\z@ \let\final@kern1\fi
      \if\final@kern1 \kern-\dimen@\fi
    \else
      \overline{\rel@kern{-0.6}\kern\dimen@#1}%
    \fi
  }%
  \macc@depth\@ne
  \let\math@bgroup\@empty \let\math@egroup\macc@set@skewchar
  \mathsurround\z@ \frozen@everymath{\mathgroup\macc@group\relax}%
  \macc@set@skewchar\relax
  \let\mathaccentV\macc@nested@a
  \if#31
    \macc@nested@a\relax111{#1}%
  \else
    \def\gobble@till@marker##1\endmarker{}%
    \futurelet\first@char\gobble@till@marker#1\endmarker
    \ifcat\noexpand\first@char A\else
      \def\first@char{}%
    \fi
    \macc@nested@a\relax111{\first@char}%
  \fi
  \endgroup
}
\makeatother

\newcommand{\ThisIsTheTitle}{TsT, $\bm{\mathsf{T\bar{T}}}$ and black strings} 
\newcommand{\ThisIsAuthorOne}{Luis Apolo,}
\newcommand{\ThisIsEmailOne}{apolo@mail.tsinghua.edu.cn}
\newcommand{\ThisIsAuthorTwo}{St\'ephane Detournay}
\newcommand{\ThisIsEmailTwo}{sdetourn@ulb.ac.be}
\newcommand{\ThisIsAuthorThree}{and Wei Song}
\newcommand{\ThisIsEmailThree}{wsong2014@mail.tsinghua.edu.cn}
\newcommand{\ThisIsTheAffiliation}{Yau Mathematical Sciences Center, Tsinghua University, Beijing 100084, China}
\newcommand{\TheseAreTheKeywords}{}

\newcommand{\ThisIsTheAbstract}{
We study the relationship between TsT transformations, marginal deformations of string theory on AdS$_3$ backgrounds, and irrelevant deformations of 2d CFTs. We show that TsT transformations of NS-NS backgrounds correspond to instantaneous deformations of the worldsheet action by the antisymmetric product of two Noether currents, holographically mirroring the definition of the $T\bar{T}$, $J\bar{T}$, $T\bar{J}$, and $J\bar{J}$ deformations of 2d CFTs. Applying a TsT transformation to string theory on BTZ $ \times\, S^3\times M^4$ we obtain a general class of rotating black string solutions, including the Horne-Horowitz and the Giveon-Itzhaki-Kutasov ones as special cases, which we show are holographically dual to thermal states in single-trace $T\bar{T}$-deformed CFTs. We also find a smooth solution interpolating between global AdS$_3$ in the IR and a linear dilaton background in the UV that is interpreted as the NS-NS ground state in the dual $T\bar{T}$-deformed CFT. This background suggests the existence of an upper bound on the deformation parameter above which the solution becomes complex. We find that the worldsheet spectrum, the thermodynamics of the black strings (in particular their Bekenstein-Hawking entropy), and the critical value of the deformation parameter match the corresponding quantities obtained from single-trace $T\bar{T}$ deformations.}

\ifjhepstyle
\title{\ThisIsTheTitle}

\author[\mbox{\fontsize{7}{8.4}{$C$}}]{\ThisIsAuthorOne}
\author[\mbox{\fontsize{7}{8.4}{$F$}}]{\ThisIsAuthorTwo}
\author[\mbox{\fontsize{7}{8.4}{$T$}}, \mbox{\fontsize{7}{8.4}{$\,C$}}]{\ThisIsAuthorThree}

\affiliation[\mbox{\fontsize{7}{8.4}{$C$}}]{\ThisIsTheAffiliation}
\affiliation[\mbox{\fontsize{7}{8.4}{$F$}}]{Physique Mathematique des Interactions Fondamentales, Universite Libre de Bruxelles and International Solvay Institutes, Campus Plaine -- CP 231, 1050 Bruxelles, Belgium}
\affiliation[\mbox{\fontsize{7}{8.4}{$T$}}]{Institute for Advanced Study, 1 Einstein Drive, Princeton, NJ 08540, USA}

\emailAdd{\ThisIsEmailOne}
\emailAdd{\ThisIsEmailTwo}
\emailAdd{\ThisIsEmailThree}

\abstract{\ThisIsTheAbstract} 

\fi

\begin{document}

\ifjhepstyle
\maketitle
\flushbottom
\fi

\long\def\symfootnote[#1]#2{\begingroup%
\def\thefootnote{\fnsymbol{footnote}}\footnote[#1]{#2}\endgroup}

\def\({\left (}
\def\){\right )}
\def\lb{\left [}
\def\rb{\right ]}
\def\lB{\left \{}
\def\rB{\right \}}

\def\Int#1#2{\int \textrm{d}^{#1} x \sqrt{|#2|}}
\def\Bra#1{\left\langle#1\right|} 
\def\Ket#1{\left|#1\right\rangle}
\def\BraKet#1#2{\left\langle#1|#2\right\rangle} 
\def\Vev#1{\left\langle#1\right\rangle}
\def\Vevm#1{\left\langle \Phi |#1| \Phi \right\rangle}\def\bbox{\bar{\Box}}
\def\til#1{\tilde{#1}}
\def\wtil#1{\widetilde{#1}}
\def\ph#1{\phantom{#1}}

\def\ra{\rightarrow}
\def\la{\leftarrow}
\def\lra{\leftrightarrow}
\def\p{\partial}
\def\barp{\bar{\partial}}
\def\diff{\mathrm{d}}

\def\sinh{\mathrm{sinh}}
\def\cosh{\mathrm{cosh}}
\def\tanh{\mathrm{tanh}}
\def\coth{\mathrm{coth}}
\def\sech{\mathrm{sech}}
\def\csch{\mathrm{csch}}

\def\a{\alpha}
\def\b{\beta}
\def\g{\gamma}
\def\d{\delta}
\def\e{\epsilon}
\def\ve{\varepsilon}
\def\k{\kappa}
\def\l{\lambda}
\def\n{\nabla}
\def\om{\omega}
\def\s{\sigma}
\def\t{\theta}
\def\z{\zeta}
\def\vp{\varphi}

\def\ss{\Sigma}
\def\dd{\Delta}
\def\GG{\Gamma}
\def\LL{\Lambda}
\def\tt{\Theta}

\def\A{{\cal A}}
\def\B{{\cal B}}
\def\C{{\cal C}}
\def\cE{{\cal E}}
\def\D{{\cal D}}
\def\F{{\cal F}}
\def\H{{\cal H}}
\def\I{{\cal I}}
\def\J{{\cal J}}
\def\K{{\cal K}}
\def\L{{\cal L}}
\def\M{{\cal M}}
\def\N{{\cal N}}
\def\O{{\cal O}}
\def\Q{{\cal Q}}
\def\P{{\cal P}}
\def\cS{{\cal S}}
\def\T{{\cal T}}
\def\W{{\cal W}}
\def\X{{\cal X}}
\def\Z{{\cal Z}}

\def\mfa{\mathfrak{a}}
\def\mfb{\mathfrak{b}}
\def\mfc{\mathfrak{c}}
\def\mfd{\mathfrak{d}}

\def\we{\wedge}
\def\re{\textrm{Re}}

\def\tilw{\tilde{w}}
\def\tile{\tilde{e}}

\def\tilL{\tilde{L}}
\def\tilJ{\tilde{J}}

\def\zz{\bar z}
\def\xx{\bar x}
\def\yy{\bar y}
\def\xp{x^{+}}
\def\xm{x^{-}}

\def\bp{\bar{\p}}
\def\note#1{{\color{red}#1}}
\def\notebf#1{{\bf\color{red}#1}}

\def\VirU1{Vir \times U(1)}
\def\VirSL2R{\mathrm{Vir}\otimes\widehat{\mathrm{SL}}(2,\mathbb{R})}
\def\U1{U(1)}
\def\u1{U(1)}
\def\SL2R{\widehat{\mathrm{SL}}(2,\mathbb{R})}
\def\sl2r{\mathrm{SL}(2,\mathbb{R})}
\def\by{\mathrm{BY}}

\def\RR{\mathbb{R}}

\def\tr{\mathrm{Tr}}
\def\bnabla{\overline{\nabla}}

\def\sint{\int_{\ss}}
\def\dsint{\int_{\p\ss}}
\def\hint{\int_{H}}

\newcommand{\eq}[1]{\begin{align}#1\end{align}}
\newcommand{\eqst}[1]{\begin{align*}#1\end{align*}}
\newcommand{\eqsp}[1]{\begin{equation}\begin{split}#1\end{split}\end{equation}}

\newcommand{\absq}[1]{{\textstyle\sqrt{\left |#1\right |}}}



\newcommand{\pp}{{\partial_+}{}}
\newcommand{\pmm}{{\partial_-}{}}


\ifprstyle
\title{\ThisIsTheTitle}

\author{\ThisIsAuthorOne}
\email{\ThisIsEmailOne}

\author{\ThisIsAuthorTwo}
\email{\ThisIsEmailTwo}

\affiliation{\ThisIsTheAffiliation}


\begin{abstract}
\ThisIsTheAbstract
\end{abstract}


\maketitle

\fi

\ifplainstyle
\begin{titlepage}
\begin{center}

\ph{.}

\vskip 4 cm

{\Large \bf \ThisIsTheTitle}

\vskip 1 cm

\renewcommand*{\thefootnote}{\fnsymbol{footnote}}

{{\ThisIsAuthorOne}\footnote{\ThisIsEmailOne} } and {\ThisIsAuthorTwo}\footnote{\ThisIsEmailTwo}

\renewcommand*{\thefootnote}{\arabic{footnote}}

\setcounter{footnote}{0}

\vskip .75 cm


\end{center}

\vskip 1.25 cm
\date{}


\end{titlepage}

\newpage

\fi

\ifplainstyle
\tableofcontents
\noindent\hrulefill
\bigskip
\fi

\section{Introduction} \label{se:introduction}

The AdS$_{d+1}$/CFT$_d$ correspondence~\cite{Maldacena:1997re} has revealed a deep relationship between bulk theories of gravity in (d+1)-dimensional Anti-de Sitter spacetimes and conformal field theories in one dimension less. It has provided many insights into both strongly coupled CFTs and quantum gravity including black holes. One among many of its manifestations is the microscopic explanation of the entropy of certain higher-dimensional black holes with an AdS$_3$ factor in the near-horizon geometry~\cite{Strominger:1996sh, Strominger:1997eq}. Recently, progress has also been achieved for asymptotically AdS black holes in higher dimensions with the advent of localization techniques~\cite{Cabo-Bizet:2018ehj, Choi:2018hmj, Benini:2018ywd}.

Several extensions of holographic dualities beyond the AdS/CFT proposal have appeared over the years. It is clear that most black holes including the real-world ones are not asymptotically AdS nor exhibit an AdS$_3$ factor in their near-horizon geometry, so AdS/CFT cannot be used as such. Various angles of attack to deal with more general black holes have been taken in recent years involving non-AdS geometries in the near-horizon region with both conformal~\cite{Guica:2008mu, Carlip:2012ff, Haco:2018ske} and non-conformal symmetries~\cite{Donnay:2015abr, Afshar:2015wjm, Afshar:2016wfy, Carlip:2017xne, Carlip:2019dbu, Grumiller:2019fmp, Aggarwal:2019iay, Grumiller:2019ygj}. Although it has been suggested that conformal symmetry could play a role in understanding generic black holes, certain non-conformal symmetries have been able to account for the microscopic entropy of the corresponding solutions~\cite{Gonzalez:2011nz, Detournay:2012pc,Barnich:2012xq, Bagchi:2012xr, Carlip:2017xne, Carlip:2019dbu} in the spirit of~\cite{Strominger:1997eq}. To understand how holography works for non-AdS spacetimes or non-conformal quantum field theories, a useful strategy is to study deformations from known examples of AdS/CFT. Interesting examples involve the so-called $\beta$~\cite{Lunin:2005jy}, noncommutative~\cite{Hashimoto:1999ut, Maldacena:1999mh} and dipole deformations~\cite{Dasgupta:2001zu, Dasgupta:2000ry, Bergman:2001rw, Alishahiha:2003ru} of $\N=4$ SYM, in which the corresponding dual gravity backgrounds are identified with so-called \emph{TsT transformations} (a succession of T-duality, shift and T-duality along $U(1) \times U(1)$ isometries) of the original geometry (for an overview see~\cite{Imeroni:2008cr}). The implications of TsT transformations have also been investigated in the D1D5 system in~\cite{ElShowk:2011cm,Song:2011sr}. 

Being able to depart from AdS spacetimes in the bulk is an important question as we stressed out above. We would like to find toy models of holographic dualities where we can compare quantities from both sides of the correspondence. In view of this, holographic dualities with $d = 2$ stand out as interesting starting points. In particular, string theory on AdS$_3$ backgrounds with pure NS-NS flux admits an exact worldsheet description in terms of an SL(2,R) WZW model that has been shown to be solvable~\cite{Maldacena:2000hw,Maldacena:2000kv,Maldacena:2001km}. The  holographic dual to this theory has been discussed in~\cite{Dijkgraaf:1998gf, Larsen:1999uk, Seiberg:1999xz, Argurio:2000tb,deBoer:1998us, Maldacena:1999bp, Gaberdiel:2007vu, Dabholkar:2007ey,Giribet:2007wp} and most recently in~\cite{Eberhardt:2018ouy, Eberhardt:2019qcl, Dei:2019osr, Eberhardt:2019ywk}. An interesting feature of WZW models is that they allow for exact conformal deformations driven by integrable marginal current-current operators~\cite{Chaudhuri:1988qb, Hassan:1992gi, Giveon:1993ph, Forste:2003km}. This allows us to reach a whole variety of new exact string theory backgrounds for which a worldsheet description is in principle available. These include in particular warped AdS$_3$ spaces, AdS$_2$~\cite{Israel:2003cx, Israel:2004vv, Israel:2004cd} and various black string solutions, in particular the Horne-Horowitz black string~\cite{Horne:1991gn, Detournay:2005fz}. These geometries have the interesting feature that they appear in the near-horizon geometry of extreme four-dimensional black holes (Kerr or Reissner-Nordstr\"om), or represent close three-dimensional analogues of asymptotically flat black holes.

In parallel, it has been recently discovered that two-dimensional CFTs admit special classes of irrelevant deformations for which the theory remains solvable. These include the $T \bar T$ deformations and their higher-spin generalizations~\cite{Zamolodchikov:2004ce,Smirnov:2016lqw,Cavaglia:2016oda}, the $J \bar T/T\bar{J}$ deformations in theories with a conserved $U(1)$ current~\cite{Guica:2017lia,Chakraborty:2018vja}, and their linear combinations~\cite{LeFloch:2019rut,Chakraborty:2019mdf,Frolov:2019xzi}. These deformations enjoy remarkable properties including closed formulae for the deformed spectrum, modular invariance ($T\bar{T}$) or covariance ($J\bar{T}/T\bar{J}$) of the torus partition functions~\cite{Dubovsky:2018bmo,Datta:2018thy,Aharony:2018bad,Aharony:2018ics,Hashimoto:2019wct}, as well as connections to strings and 2d gravity~\cite{Dubovsky:2012wk,Cavaglia:2016oda,Dubovsky:2017cnj,Cardy:2018sdv,Baggio:2018gct,Frolov:2019nrr,Aguilera-Damia:2019tpe,Callebaut:2019omt,Anous:2019osb,Tolley:2019nmm}. The holographic interpretation of these irrelevant deformations include the double-trace version which changes the boundary conditions but does not change the local bulk geometry~\cite{McGough:2016lol,Bzowski:2018pcy,Guica:2019nzm}, as well as a single-trace version which also changes the local geometry~\cite{Giveon:2017nie,Chakraborty:2018vja,Apolo:2018qpq}. In particular, the single-trace $T\bar T$ deformation is dual to a geometry that interpolates between AdS$_3$ and a linear dilaton background that is relevant in the study of Little String Theory~\cite{Giveon:2017nie}; while the single-trace $J\bar T/T\bar{J}$ deformation is holographically dual to the warped AdS$_3$ geometry that is relevant in the study of extremal black holes~\cite{Chakraborty:2018vja,Apolo:2018qpq,Apolo:2019yfj}. For additional references on irrelevant deformations see e.g.~\cite{Bonelli:2018kik,Chakraborty:2018kpr,Taylor:2018xcy,Donnelly:2018bef,Chen:2018eqk,Hartman:2018tkw,Gorbenko:2018oov,Gross:2019ach,Asrat:2019end} and the reviews~\cite{Giveon:2019fgr,Jiang:2019hxb}.

In this paper we investigate the relationship between marginal deformations of string theory on AdS$_3$, deformations of the AdS$_3$ geometry, and the corresponding deformations of the dual CFT. TsT transformations play a central role in this story, which is analogous to the case of non-commutative, dipole, and $\b$ deformations, but with purely NS-NS backgrounds in the bulk. In particular, we find that TsT transformations of NS-NS backgrounds parametrized by the dimensionless constant $\l$ can be described by the following deformation of the bosonic part of the worldsheet action $S_{\textrm{WS}}$,   
 \eq{
  \frac{\p S_{\textrm{WS}}}{\p {\l}} = -\frac{1}{\pi k } \int  {\bm j}_{(1)} \wedge \bm{j}_{({\bar 2})}.
  \label{tstdeformationintro}
  }
In eq.~\eqref{tstdeformationintro}, $k$ is the number of NS5 branes generating the undeformed background, the subscripts $(1)$ and $({\bar 2})$ correspond to the two target space coordinates $X^1$ and $X^{{\bar 2}}$ along which the TsT transformation is performed, while ${\bm j}_{(1)}$ and ${\bm j}_{({\bar 2})}$ denote the worldsheet Noether one-forms generating shifts along these coordinates. 

We conjecture that string theory on TsT-transformed AdS$_3\times S^3 \times M^4$ backgrounds with NS-NS flux ($M^4$ is a compact four-dimensional manifold), whose worldsheet action satisfies eq.~\eqref{tstdeformationintro}, is holographically dual to a single-trace deformation of the dual CFT. Before the deformation, the long string sector of string theory on AdS$_3\times S^3 \times M^4$ is dual to a symmetric product  $\M^p /S_p$ where $\M$ is a CFT with central charge $6k$ and $p$ denotes the number of NS1 strings generating the background~\cite{Argurio:2000tb,Giveon:2005mi,Eberhardt:2019qcl}. The single trace deformation corresponds to a deformation of the seed theory $\M \to \M_\mu$ such that the deformed theory remains a symmetric product, namely $(\M_\mu)^p/S_p$.\footnote{The single-trace deformation of the symmetric product ${\cal M}^p/S_p$ was first proposed in~\cite{Giveon:2017nie}. For comparison, the double trace deformation is defined by $\p_\mu S^{tot}_{\mu} = -\frac{1}{\pi} \int J_{(1)}^{tot} \wedge J_{({\bar 2})}^{tot} $ where $S^{tot}_{\mu}$ is the action of the deformed theory and $J^{tot}_{(m)}$ is a Noether current in the entire deformed theory, in contrast to just the seed~\eqref{cftdeformationintro}.} More explicitly, the action of the deformed seed theory ${\cal M}_{\mu}$ is defined by
   \eq{
    \frac{\p S^i_{\mu}}{\p \mu} = -\frac{1}{\pi } \int J_{(1)}^i \wedge J_{({\bar 2})}^i, \label{cftdeformationintro}
      }
where $\mu$ is the deformation parameter. In eq.~\eqref{cftdeformationintro}, $J^i_{(1)}$ and $J^i_{({\bar 2})}$ denote the instantaneous Noether one-forms generating translations along the $X^1$ and $X^{{\bar 2}}$ coordinates in the $i^{\textrm{th}}$ copy of the deformed seed ${\cal M}_\mu$. In order to obtain locally non-AdS$_3$ geometries at least one of the TsT coordinates must belong to the AdS$_3$ part of the AdS$_3\times S^3 \times M^4$ background. Correspondingly, the holographic dual is expected to be described by an irrelevant deformation of the dual CFT. In particular,  when both of the TsT coordinates are chosen from AdS$_3$, we will show that the irrelevant deformation corresponds to the single-trace $T\bar{T}$ deformation. This is a finite temperature generalization of the proposal made in~\cite{Giveon:2017nie}, the latter of which is supported by matching the string theory spectrum on the Ramond vacuum. On the other hand, when only one of the $U(1)$ directions belongs to the AdS$_3$ factor, the holographic dual is described by the single-trace $J\bar T$/$T\bar J$ deformation~\cite{Apolo:2019yfj}, a conjecture supported by matching the string theory spectrum at both zero~\cite{Apolo:2018qpq,Chakraborty:2019mdf} and finite temperature~\cite{Apolo:2019yfj}, and reproducing the thermodynamics of the bulk from the boundary side of the correspondence~\cite{Apolo:2019yfj}. Finally, if both of the TsT directions are taken along $U(1)$ isometries of the internal manifold then the deformation of the dual CFT is a marginal $J\bar{J}$ deformation. 

In this paper we provide further evidence for the correspondence between TsT transformations of AdS$_3$ backgrounds  \eqref{tstdeformationintro} and irrelevant deformations of the dual CFT \eqref{cftdeformationintro}. Applying a TsT transformation to string theory on BTZ$\,\times S^3 \times M^4$ backgrounds  we obtain finite temperature spacetimes that are dual to thermal states in single-trace $T\bar{T}$-deformed CFTs. In particular, we show that
  \begin{itemize}
\item[(\emph{i})] the (long) string spectrum on the TsT-transformed background matches the spectrum of single-trace $T\bar{T}$-deformed CFTs. 
\item[(\emph{ii})] The thermodynamics of the resulting black string solutions match the thermodynamics of $T\bar{T}$ deformations. In particular, we reproduce the Bekenstein-Hawking entropy of black strings from the density of states of the dual field theory.
\item[(\emph{iii})] There is a critical value of the deformation parameter ${\l}_c$ such that, for ${\l} \le {\l}_c$, the ground state of the dual $T\bar{T}$-deformed CFT has a real energy and corresponds to a smooth solution in the bulk. On the other hand, for ${\l} > {\l}_c$, the ground state energy becomes complex and, correspondingly, the bulk solution becomes complex as well. 
\item[(\emph{iv})] The TsT background describes other known black string solutions in special cases. In particular, the Horne-Horowitz black string~\cite{Horne:1991gn} corresponds to the non-rotating TsT background with $\lambda={1/ 2}$ after a gauge transformation. Similarly, the black string of~\cite{Giveon:2017nie} corresponds to the non-rotating TsT black string with $\l = 1/2$ after a shift of the dilaton and a gauge transformation.
\end{itemize}

The paper is organized as follows. In Section~\ref{se:tst} we relate TsT transformations to marginal deformations of the worldsheet action, derive general expressions for the deformed spectrum, and discuss the relationship between TsT transformations and solvable irrelevant deformations. In Section~\ref{se:tstbtz} we perform a TsT transformation on the BTZ $\times S^3 \times M^4$ background, match the string theory spectrum to that of $T\bar{T}$-deformed CFTs, and discuss the relationship between the TsT transformation and current-current deformations of the worldsheet. We compute the charges, entropy, and thermodynamic potentials of the TsT black string in Section~\ref{se:thermodynamics}. Therein, we also show that these quantities match the corresponding expressions derived in single-trace $T\bar{T}$-deformed CFTs. In Section~\ref{se:phasespace} we discuss the phase space of TsT-transformed solutions including the ground state, the $T\bar{T}$ flow of states, and the superluminal $\l < 0$ sign of the deformation. In Section~\ref{se:hhgik} we describe the relationship between the TsT black string and the Horne-Horowitz and Giveon-Itzhaki-Kutasov backgrounds. In Appendix~\ref{ap:charges} we review the covariant formulation of gravitational charges while in Appendix~\ref{ap:tstbackground} we describe alternative parametrizations of the black string.

\section{TsT and irrelevant deformations} \label{se:tst}

In this section we consider generic TsT transformations of IIB string theory with NS-NS background fields. We show that these TsT transformations are equivalent to instantaneous deformations of the worldsheet action by the antisymmetric product of two Noether currents. These deformations mirror the definition of the $T\bar{T}$ and $J\bar{T}/T\bar{J}$ deformations of 2d CFTs and we propose a precise correspondence with TsT transformations of AdS$_3 \times S^3 \times M^4$ backgrounds. Finally, provided that certain assumptions hold, we derive general expressions for the spectrum of strings winding on TsT-transformed backgrounds.


\subsection{TsT as a worldsheet deformation}

Let us begin by considering the bosonic sector of IIB string theory on an arbitrary background supported by NS-NS flux. The worldsheet action is given by
  \eq{
  \til{S} =- \frac{1}{4\pi \ell_s^{2}} \int d^2z \big( \sqrt{-\eta} \eta^{ab} \til{G}_{\mu\nu} + \e^{ab} \til{B}_{\mu \nu} \big)  \p_a \til{X}^{\mu} \p_b \til{X}^{\nu}  = \frac{1}{ 2\pi\ell_s^2} \int d^2z\,  \til{M}_{\mu\nu} \p \til{X}^{\mu} \bp \til{X}^{\nu}, \label{hataction}
  }
where $\ell_s$ is the string scale, $z = \tau + \s$ and $\bar{z} = \tau - \s$ are the worldsheet coordinates with $\p \equiv \p_z$ and $\bp \equiv \p_{\bar{z}}$, $\til{M}_{\mu\nu} = \til{G}_{\mu\nu} + \til{B}_{\mu\nu}$, and the worldsheet metric $\eta_{ab}$ and Levi-Civita symbol $\epsilon_{ab}$ satisfy
  \eq{
  \eta_{z\bar{z}} = - 1/2,  \qquad \eta^{z\bar{z}} = -2, \qquad  \e_{z\bar{z}} = - \e^{z\bar{z}} = 1.
  }
We assume that the background fields $\til{G}_{\mu\nu}$, $\til{B}_{\mu\nu}$, and the dilaton $\til{\Phi}$ are independent of the target space coordinates $\til{X}^n$ with $n = 1, \bar{2}$ such that the action is invariant under two $U(1)$ symmetries generated by\footnote{The motivation behind the $m = 1$ and $\bar{m} = \bar{2}$ indices in $X^m$ and $X^{\bar{m}}$ will be explained in Section~\ref{se:tstspectrum} and is related to the chiral properties of the Noether currents generating translation along these coordinates.}
  \eq{
  \xi_{(n)} = \p_{\til{X}^n}, \qquad n = 1, \bar{2}.
  }
In particular, the Noether currents associated with the $\xi_{(n)}$ Killing vectors read
  \eq{
  \til{j}^a_{(n)} &= - \ell_s^{-2} \big(\sqrt{-\eta} \eta^{ab} \til{G}_{n \mu} \p_b \til{X}^{\mu} + \e^{ab} \til{B}_{n \mu} \p_b \til{X}^{\mu} \big).   \label{noethercurrents}
  }
These currents are generically not chiral and their conservation law corresponds to the equations of motion of the $\til{X}^n$ coordinates. Alternative expressions for the Noether currents that are useful in what follows include
    \eq{ 
     \til{j}_{(n)}& \equiv \til{j}^a_{(n)} \p_a = \ell_s^{-2} \big( \til{M}_{n\mu} \bp X^{\mu} \p + \til{M}_{\mu n} \p X^{\mu} \bp \big),\label{noethervector} \\
       \til{{\bm j}}_{(n)}&\equiv 2 \til{j}_{(n)}^a \eta_{ab} d z^b =  -\ell_s^{-2} \big(\til{G}_{n\mu} d \til{X}^{\mu}  - \til{B}_{n\mu } \star d \til{X}^{\mu}\big) \label{noetherform},
  }
where the bold notation serves to distinguish the vector $\til{j}_{(n)}$ from the one-form $\til{{\bm j}}_{(n)}$ representations of the current.\footnote{Note that the Noether current~\eqref{noethercurrents} is actually a vector density and that we have included an additional factor of 2 in the definition of the one-form~\eqref{noetherform}.}

We now perform a TsT transformation of the background fields that consists of T-duality along $\til{X}^1$, followed by a shift $\til{X}^{\bar{2}}  = X^{\bar{2}} - 2 \til{\l}  X^1$ with $\til{X}^1 = X^1$, and another T-duality along the $X^1$ coordinate. The TsT transformation is characterized by a dimensionless parameter $\til{\l}$. After the TsT transformation, the transformed metric $G_{\mu\nu}$, $B_{\mu\nu}$ field, and dilaton $\Phi$ can be obtained from the original fields via (see for example~\cite{Lunin:2005jy,CatalOzer:2005mr})
  \eq{
  M = \til{M} \bigg (I+{2 \til{\l}\over \ell_s^2} \GG\til{M} \bigg)^{-1}, \qquad \Phi_{\textrm{TsT}} = \til{\Phi} + \frac{1}{4} \log \bigg( \frac{\textrm{det } G_{\mu\nu}}{\textrm{det } \til{G}_{\mu\nu}} \bigg), \label{tst} 
  }
where $\GG$ is defined by
  \eq{
  \GG_{\mu\nu} = \d^{1}_{\mu} \d^{{\bar{2}}}_{\nu} - \d^{{\bar{2}}}_{\mu} \d^{1}_{\nu}. \label{gammadef}
  }
Note that when the $\til{j}_{(1)}$ and $\til{j}_{({\bar{2}})}$ currents are both chiral or both antichiral, $\til{M}$ has two rows or two columns of zeros. As a result,  $\Gamma \til{M}=0$ and the TsT transformation is trivial. In this paper, we will only consider TsT transformations that lead to different worldsheet theories. 

After the TsT transformation the worldsheet action is given by
  \eq{
  S = {1\over 2\pi \ell_s^{2}} \int d^2z\, {M}_{\mu\nu} \p {X}^{\mu} \bp {X}^{\nu}, \label{tstaction}
  }
where $X^{\mu}$ denote the target space coordinates after TsT. From eq.~\eqref{tst}, it is not difficult to see that the background field $M_{\mu\nu}$ obeys the differential equation
  \eq{
  \frac{\p M}{\p{\til{\l}}}  = -2\ell_s^{-2} M  \GG M.  \label{tstidentity}
  }
We note that the Noether currents $j_{(n)}$ of the deformed theory, which are given by eq.~\eqref{noethercurrents} with $\til{M}_{\mu\nu} \to {M}_{\mu\nu}$ and $\til{X}^{\mu} \to {X}^{\mu}$, also satisfy
  \eq{
  & {\bm j}_{(1)} \wedge  {\bm j}_{({\bar{2}})}  = \frac{1}{\ell_s^4} (M \GG M)_{\mu\nu} \p X^{\mu} \bp X^{\nu} \, dz \we d \bar{z}. \label{tstidentity2}
  }

Thus, eqs.~\eqref{tstidentity} and~\eqref{tstidentity2} imply that the worldsheet action satisfies the following differential equation
  \eq{
  \frac{\p S}{\p \til{\l}} = -\frac{1}{\pi} \int  {\bm j}_{(1)} \wedge \bm{j}_{({\bar{2}})} . \label{tstdeformation}
  }
We conclude that TsT transformations correspond to instantaneous deformations of the worldsheet action by the antisymmetric product of two Noether currents. This is a universal result that applies to the action of strings on any TsT-transformed background supported by NS-NS flux, generalizing the observations made in~\cite{Apolo:2019yfj}.\footnote{The generalization of our results to backgrounds supported by R-R flux may be possible using the hybrid formalism~\cite{Berkovits:1999im}.} Note that the TsT-origin of eq.~\eqref{tstdeformation} guarantees that the deformation is exactly marginal to leading order in $\alpha' \propto \ell_s^2$. This suggests that the analysis of ref.~\cite{Chaudhuri:1988qb}, which showed that chiral-antichiral current-current deformations are exactly marginal, can be extended to the antisymmetric product of two Noether currents. Finally, we note that the worldsheet deformation~\eqref{tstdeformation} mirrors the definition of the $T\bar{T}$ and $J\bar{T}/T\bar{J}$ deformations of 2d CFTs which are also instantaneous deformations by the antisymmetric product of two conserved currents~\cite{Smirnov:2016lqw,Cavaglia:2016oda,Guica:2017lia}. The connection between TsT deformations and irrelevant deformations is discussed in detail in Section~\ref{se:tstirrelevant}. 

\subsection{The worldsheet spectrum} \label{se:tstspectrum}

We now derive general expressions for the spectrum of strings winding on TsT-transformed backgrounds that are valid provided that certain conditions, to be described momentarily, hold. An important feature of TsT transformations is that the equations of motion and Virasoro constraints of the deformed theory~\eqref{tstaction} can be obtained from those of the undeformed one by the following change of coordinates~\cite{Frolov:2005dj,Rashkov:2005mi,Alday:2005ww}\footnote{The equivalence of the equations of motion and Virasoro constraints of the deformed and undeformed theories is most easily seen by writing the change of coordinates~\eqref{xshift} in matrix notation such that $\p \hat{X} = \p {X} \cdot M \cdot \hat{M}^{-1} $ and $\bp \hat{X} = \hat{M}^{-1} \cdot M \cdot \bp X$ where $X$ and $\hat{X}$ denote the (un)deformed target space coordinates.\label{matrixrep}}
  \eq{
  \p_a \hat{X}^n&=\p_a X^n - 2\til{\l} \, \e^{nn'} \e_{ab} j^{b}_{(n')}, \qquad \e^{1{\bar{2}}} = -1, \quad n,n'\in \{1,{\bar{2}}\}, \label{xshift}
  }
where we sum over repeated indices and labels. Note that eq.~\eqref{xshift} implies that the $\hat{X}^{n}$ coordinates satisfy nonlocal boundary conditions, which distinguish the latter from the undeformed $\til{X}^n$ coordinates introduced earlier. In particular, since the background fields are assumed to be independent of the $\til{X}^n$ coordinates, we have
  \eq{
  \hat{G}_{\mu\nu} = \til{G}_{\mu\nu}, \qquad \hat{B}_{\mu\nu} = \til{B}_{\mu\nu}, \qquad \hat{\Phi} = \til{\Phi}. \label{tildehat}
  }
 The twisted boundary conditions satisfied by the $\hat{X}^n$ coordinates can be deduced by integrating $\oint (\p - \bp)\hat{X}^{n}$ with $\oint \equiv \int^{2\pi}_0 d\s$ such that
    \eq{
    \hat{X}^n(\s + 2\pi) = \hat{X}^n(\s) + 2\pi \g^{(n)}, \label{twistedbc}
    }
where the $\g^{(n)}$ parameters are given by
  \eq{
  \g^{(n)} =  \frac{1}{2\pi} \oint \p_\s X^n - 2\til{\l} \, \e^{nn'} p_{(n')}. \label{defgamma}
  }
In eq.~\eqref{defgamma} $p_{(n)}$ denotes the physical momenta of the string along the $X^n$ coordinate, i.e.~the momenta of the string \emph{after} the TsT transformation, which is defined by
  \eq{
  p_{(n)} = \frac{1}{2\pi} \oint j^{\tau}_{(n)} = \frac{1}{2\pi} \oint  j^z_{(n)} + j^{\bar{z}}_{(n)}. \label{momentumdef}
  }
Note that the total derivative term in eq.~\eqref{defgamma} vanishes unless $X^{n}$ is compact, in which case the string can wind $|w^{(n)}|$ times along the $X^{n}$ coordinate
  \eq{
  \frac{1}{2\pi} \oint \p_\s X^n = w^{(n)}, \qquad w^{(n)} \in \mathbb{Z}.
  }
For completeness, we assume arbitrary winding parameters $w^{(n)}$ along the TsT coordinates.

The twisted boundary conditions~\eqref{twistedbc} relate the spectrum of strings on the TsT-transformed background to the spectrum of strings before the TsT transformation. The spectrum is sensitive to the directions along which the strings are allowed to wind, and these may not coincide with the TsT coordinates. For example, this is the case in the holographic description of single-trace $J\bar{T}/T\bar{J}$ deformations where strings wind only along one of the TsT coordinates. For this reason, we will assume that the background features additional $U(1)$ isometries. In order to derive general expressions for the worldsheet spectrum it is also useful to introduce two kinds of indices, $X^m$ and $X^{\bar{m}}$, that are distinguished by the chiral properties of the Noether currents generating translations along these coordinates. Therefore, we will assume that the background $\hat{G}_{\mu\nu}$, $\hat{B}_{\mu\nu}$, and $\hat{\Phi}$ fields are independent of the following coordinates
  \eq{
  \hat{X}^{m}, \qquad \hat{X}^{\bar{m}}, \qquad m, \bar{m} \in \{1, \bar{2}, \dots N \}. \label{newU1coords}
  }  %
The boundary conditions satisfied by these coordinates are parametrized by
  \eq{
  \hat{X}^m(\s + 2\pi) = \hat{X}^m (\s) + 2 \pi \g^{(m)}, \qquad   \hat{X}^{\bar{m}} (\s + 2\pi) = \hat{X}^{\bar{m}} (\s) + 2 \pi \g^{(\bar{m})}, \label{newbc}
  }
while the rest of the target space coordinates $\hat{X}^{i}$, where $i\in\{N+1 , \dots, 10\}$, obey trivial boundary conditions. In particular, we will continue to use $m=1$ and $\bar{m} = \bar{2}$ to denote the TsT directions such that $\g^{(1)}$ and $\g^{(\bar{2})}$ are momentum-dependent and given by eq.~\eqref{defgamma}. In contrast, the other $\g^{(m)}$ and $\g^{(\bar{m})}$ parameters satisfy
  \eq{
  \g^{(m)} = w^{(m)}, \qquad \g^{(\bar{m})} = w^{(\bar{m})}, \qquad m, \bar{m} \in \{3, \bar{4}, \dots N \}, \label{defothergamma}
  }
where $|w^{(m)}|$ and $|w^{(\bar{m})}|$ are integers that count the number of times the string winds along the corresponding $X^m$ and $X^{\bar{m}}$ directions.

We can implement the boundary conditions~\eqref{newbc} by a shift of coordinates, 
  \eq{
  \hat{X}^m = \til{X}^m + \g^{(m)} z, \qquad   \hat{X}^{\bar{m}} = \til{X}^{\bar{m}} - \g^{(\bar{m})} \bar{z}, \label{TsTshift}
  }
where $\til{X}^{\mu}$ satisfy trivial boundary conditions and describe the target space coordinates of the string before the TsT transformation (in contrast, the $\hat{X}^{\mu}$ coordinates satisfy twisted and winding boundary conditions and describe the string after TsT). Eq.~\eqref{TsTshift}  remains a solution to the equations of motion provided that the background fields satisfy
  \eq{
  \p_{[\mu} \hat{M}_{|m| \nu]} = 0, \qquad \p_{[\mu} \hat{M}_{\nu] \bar{m}} = 0,\label{Mconstraint}
  }
which implies that $\hat{M}_{mm'}$, $\hat{M}_{\bar{m}\bar{m}'}$, and $\hat{M}_{m\bar{m}}$ are constant for all $m$ and $\bar{m}$.  Eq.~\eqref{Mconstraint} is the main assumption needed in our derivation of the spectrum. It constrains the background field $\hat{M}_{\mu\nu}$, the two TsT coordinates, as well as the winding directions along which we can impose the boundary conditions~\eqref{newbc}. Note that the constraint~\eqref{Mconstraint} is satisfied, in particular, by the AdS$_3 \times S^3  \times M^4$ background relevant in the holographic description of $T\bar{T}$ and $J\bar{T}/T\bar{J}$-deformed CFTs. 

Interestingly, the constraint~\eqref{Mconstraint} implies the existence of chiral currents for which the shift~\eqref{TsTshift} may be interpreted as a spectral flow transformation. These chiral currents are given by
  \eq{
  \hat{h}_{(m)} &= 2 \ell_s^{-2} \hat{G}_{m \mu } \p \hat{X}^{\mu} \bp = \hat{j}_{(m)} - \hat{M}_{m\mu}  \big( \bp \hat{X}^{\mu}\p - \p \hat{X}^{\mu} \bp \big), \label{hdef} \\
    \hat{h}_{(\bar{m})} &= 2 \ell_s^{-2} \hat{G}_{\bar{m} \mu} \bp \hat{X}^{\mu} \p = \hat{j}_{(\bar{m})} + \hat{M}_{\mu\bar{m}}  \big( \bp \hat{X}^{\mu}\p - \p \hat{X}^{\mu} \bp \big),  \label{hbardef}
  }
where $\hat{j}_{(m)}$ and $\hat{j}_{(\bar{m})}$ are the Noether currents associated with translations along $\hat{X}^m$ and $\hat{X}^{\bar{m}}$, the latter of which are given by eq.~\eqref{noethercurrents} with $\til{X}^{\mu} \to \hat{X}^{\mu}$ (recall that $\til{M}_{\mu\nu} = \hat{M}_{\mu\nu}$). The other terms on the right-hand side of eqs.~\eqref{hdef} and~\eqref{hbardef} are topological in the sense that they are conserved identically (off-shell) by means of eq.~\eqref{Mconstraint}. Thus, conservation of the Noether currents implies that $\hat{h}_{(m)}$ and $\hat{h}_{(\bar{m})}$ are chirally conserved, 
  \eq{
  \bp \hat{h}^{\bar{z}}_{(m)} = 0, \qquad \p \hat{h}^{z}_{(\bar{m})} = 0, \label{chiralconservation}
  }
which motivates the introduction of the $m$ and $\bar{m}$ indices. The constraint~\eqref{Mconstraint} is responsible for turning the global isometries of the background into infinite-dimensional sets of local symmetries on the worldsheet theory. Note that eq.~\eqref{chiralconservation} implies that the Noether currents associated with TsT transformations are chiral and antichiral up to topological terms. As a result, the instantaneous deformation~\eqref{tstdeformation} is closely related to the current-current deformations considered in ref.~\cite{Chaudhuri:1988qb}.  Finally, we note that the $\hat{h}_{(m)}$ and $\hat{h}_{(\bar{m})}$ currents shift by a constant under the change of coordinates~\eqref{TsTshift}, suggesting that the latter can be interpreted as a spectral flow transformation, an observation that can be verified explicitly for AdS$_3 \times S^3 \times M^4$ backgrounds.

The shift of coordinates introduced in eq.~\eqref{TsTshift} relates the components of the stress tensor ${T}_{ab} = \frac{4\pi}{\sqrt{-\eta}} \frac{\d {S}}{\d \eta^{ab}}$ before and after the TsT transformation. In terms of the hatted variables, i.e.~after TsT, the components of the stress tensor read
  \eq{
  \hat{T} \equiv \hat{T}_{zz} = - \ell_s^{-2} \hat{G}_{\mu\nu} \p \hat{X}^{\mu} \p \hat{X}^{\nu}, \qquad  \hat{\bar{T}} \equiv \hat{\bar{T}}_{zz} = - \ell_s^{-2}\hat{G}_{\mu\nu} \bp \hat{X}^{\mu} \bp \hat{X}^{\nu}. \label{hatTTbar}
  }
Using the change of coordinates~\eqref{TsTshift} we then find
  \eq{
  \hat{T} & =  \til{T} - \hat{h}^{\bar{z}}_{(m)} \gamma^{(m)} + \ell_s^{-2}\hat{G}_{mm'} \g^{(m)} \g^{(m')},   \label{hatT} \\
  \hat{\bar{T}} &= \til{\bar{T}} + \hat{h}^z_{(\bar{m})} \g^{(\bar{m})} + \ell_s^{-2} \hat{G}_{\bar{m}\bar{m}'} \g^{(\bar{m})} \g^{(\bar{m}')}, \label{hatTbar}
  }
where $\til{T} = - \ell_s^{-2}\til{G}_{\mu\nu} \p \til{X}^{\mu} \p \til{X}^{\nu}$ and $\til{\bar{T}} = - \ell_s^{-2} \til{G}_{\mu\nu} \bp \til{X}^{\mu} \bp \til{X}^{\nu}$ denote the components of the stress tensor before the TsT transformation. In the derivation of eqs.~\eqref{hatT} and~\eqref{hatTbar} we have used eq.~\eqref{tildehat} and the definition of the chiral currents given in eqs.~\eqref{hdef} and~\eqref{hbardef}.

The zero modes of the chiral currents $\hat{h}_{(m)}$ and $\hat{h}_{(\bar{m})}$ are related to the physical momenta of the string on the TsT-transformed background $M_{\mu\nu}$. In order to see this we note that the nonlocal change of coordinates~\eqref{xshift} implies that~\cite{Alday:2005ww}\footnote{Eq.~\eqref{currentmap} can be easily proven using the matrix representation of eq.~\eqref{xshift} given in footnote~\ref{matrixrep}.} 
  \eq{
  \hat{j}_{(m)}(\hat{X}) = j_{(m)}(X), \qquad \hat{j}_{(\bar{m})} (\hat{X})= j_{(\bar{m})}(X), \label{currentmap}
  }
where $j_{(m)}$ and $j_{(\bar{m})}$ are the Noether currents of the deformed theory. Thus, using the boundary conditions~\eqref{newbc} and the definition of the momenta given in eq.~\eqref{momentumdef} we find
  \eq{
  \frac{\ell_s^2}{2\pi} \oint h^{\bar{z}}_{(m)} &= \ell_s^2 p_{(m)} +  \hat{M}_{mm'} \g^{(m')} + \hat{M}_{m\bar{m}} \g^{(\bar{m})} + \frac{1}{2\pi} \oint \hat{M}_{m i} \p_{\s} \hat{X}^i,  \\
   \frac{\ell_s^2}{2\pi} \oint h^{{z}}_{(\bar{m})} &= \ell_s^2 p_{(\bar{m})} - \hat{M}_{\bar{m}' \bar{m}} \g^{(\bar{m}')} - \hat{M}_{m\bar{m}} \g^{({m})} - \frac{1}{2\pi} \oint \hat{M}_{i \bar{m}} \p_{\s} \hat{X}^i.
  }
Consequently, the zero modes of the stress tensor $\hat{L}_0 \equiv - \frac{1}{2\pi} \oint \hat{T}$ and $\hat{\bar{L}}_0 \equiv - \frac{1}{2\pi} \oint \hat{\bar{T}}$ satisfy
  \eq{
  \hat{L}_0 &= \til{L}_0 + p_{(m)} w^{(m)} + 2 \til{\l}  p_{(1)} p_{(\bar{2})} +\ell_s^{-2} \hat{M}_{m\bar{m}} \g^{(m)} \g^{(\bar{m})} + \frac{\g^{(m)}}{2\pi\ell_s^2} \oint \hat{M}_{mi}  \p_{\s} \hat{X}^{i},  \label{tstL0} \\
   \hat{\bar{L}}_0 &= \til{\bar{L}}_0 -  p_{(\bar{m})} w^{(\bar{m})} + 2 \til{\l}  p_{(1)} p_{(\bar{2})} + \ell_s^{-2}\hat{M}_{m\bar{m}} \g^{(m)} \g^{(\bar{m})} + \frac{\g^{(\bar{m})}}{2\pi\ell_s^{2}} \oint \hat{M}_{i\bar{m}}  \p_{\s} \hat{X}^{i}, \label{tstL0bar}
  }
 where $\til{L}_0 \equiv - \frac{1}{2\pi} \oint \til{T}$ and $\til{\bar{L}}_0 \equiv - \frac{1}{2\pi} \oint \til{\bar{T}}$ denote the zero modes of the stress tensor before the TsT transformation and we used the definitions of the $\g^{(m)}$ and $\g^{(\bar{m})}$ parameters given in eqs.~\eqref{defgamma} and~\eqref{defothergamma}. 
 
 Eqs.~\eqref{tstL0} and~\eqref{tstL0bar} relate the spectrum of strings winding on the TsT-transformed background $(\hat{L}_0, \hat{\bar{L}}_0)$ to the spectrum of strings before the TsT transformation $(\til{L}_0, \til{\bar{L}}_0)$. These equations are valid for generic TsT backgrounds provided that we choose one TsT direction along $\hat{X}^m$ and another one along $\hat{X}^{\bar{m}}$ such that the original background field $\hat{M}_{\mu\nu}$ satisfies eq.~\eqref{Mconstraint}. If we impose further constraints on the background fields such as
  \eq{
  \p_{\mu} \hat{M}_{mi} =  \p_{\mu} \hat{M}_{i\bar{m}} =0 , \qquad \hat{M}_{m\bar{m}} = 0, \label{Mconstraint2}
  }
then the worldsheet spectrum simplifies to
  \eq{
  \hat{L}_0 &= \til{L}_0 +  p_{(m)} w^{(m)} + 2 \til{\l}  p_{(1)} p_{(\bar{2})},  \label{tstL02} \\
   \hat{\bar{L}}_0 &= \til{\bar{L}}_0 - p_{(\bar{m})} w^{(\bar{m})} + 2 \til{\l}  p_{(1)} p_{(\bar{2})}.\label{tstL0bar2}
  }
  The constraint~\eqref{Mconstraint2} may seem to be too restrictive but it turns out to be satisfied by all locally AdS$_3 \times S^3  \times M^4$ backgrounds provided that the TsT and winding directions are chosen appropriately. In particular, this background is relevant in the holographic description of the $T\bar{T}$ and $J\bar{T}/T\bar{J}$ deformations as we describe next.
  

\subsection{Marginal vs irrelevant deformations} \label{se:tstirrelevant}
Let us now consider string theory on
  \eq{
  \textrm{AdS}_3 \times S^3 \times M^4, \label{AdS3S3M4}
  }
where the AdS$_3$ part of the background may correspond to the NS vacuum, the massless BTZ black hole (Ramond vacuum), a general BTZ black hole or any other orbifold of AdS$_3$ that preserves two $U(1)$ isometries. The worldsheet action on these backgrounds is described by a WZW model at level $k = \ell^2/\ell_s^2$ where $\ell$ and $\ell_s$ denote the AdS and string scales, respectively. There is evidence that the long string sector of string theory on AdS$_3 \times S^3 \times M^4$ supported by NS-NS flux is dual to a symmetric product orbifold~\cite{Argurio:2000tb,Giveon:2005mi,Eberhardt:2019qcl}
  \eq{
   {\cal M}^p/S_p,  \label{symprod}
  }
where ${\cal M}$ denotes a CFT with central charge $c_\M = 6k$ that depends on the choice of the compact $M^4$ manifold while $k$ is the number of NS5 branes and $p$ is the number of long strings generating the background. The structure of the dual CFT allows us to define both single-trace and double-trace bilinear deformations of the action. In particular, the single-trace $T\bar{T}$ and $J\bar{T}/T\bar{J}$ deformations of the dual CFT are instantaneous deformations of the seed theory ${\cal M} \to {\cal M}_\mu$ defined by
  \eq{
  \frac{\p S^i_{\mu}}{\p \mu} = - \frac{1}{\pi} \int J^i_{(1)} \wedge J^i_{(\bar{2})}, \label{cftdeformation}
  }
where $\mu$ denotes the deformation parameter in the field theory, $S^i_\mu$ is the action in the $i^{\textrm{th}}$ copy of the deformed seed ${\cal M}_\mu$, and $J^i_{(1)}$, $J^i_{(\bar{2})}$ are Noether one-forms of ${\cal M}_\mu$ that we define shortly. While the full holographic description of string theory on AdS$_3$ is not yet known, we will assume that single-trace deformations analogous to eq.~\eqref{cftdeformation} can still be defined such that the deformed theory is described by the symmetric product $({\cal M}_\mu)^p/S_p$ (see~\cite{Chakraborty:2019mdf} for a related discussion).

There is a striking similarity between the marginal deformation of the worldsheet action~\eqref{tstdeformation} and the irrelevant deformation of the dual CFT~\eqref{cftdeformation}. Letting ${\bm j}_{(m)}$ denote the Noether one-forms generating translations along the TsT directions on the worldsheet, and $J_{(m)}$ the corresponding one-forms in the dual field theory,  we propose that
\begin{center}
\noindent\parbox{13.5cm}{
\emph{string theories defined by the worldsheet action~\eqref{tstdeformation} which describes NS-NS background fields obtained by a TsT transformation of AdS$_3 \times S^3 \times M^4$ are holographically dual to single-trace deformations of a CFT of the form~\eqref{cftdeformation}.}
}
\end{center}
 To obtain a non-AdS$_3$ geometry, at least one of the $U(1)$ directions characterizing the TsT transformation must belong to the AdS$_3$ factor. Consequently, the holographic dual is expected to be described by an irrelevant deformation of a CFT. In particular, when both of the TsT directions are taken along translational $U(1)$ isometries of AdS$_3$, we will show that the irrelevant deformation corresponds to the single-trace $T\bar T$ deformation. This observation is supported by matching the string theory spectrum on the Ramond vacuum~\cite{Giveon:2017myj} and --- as we will show in this paper --- matching the string theory spectrum on finite-temperature backgrounds and reproducing the thermodynamics of TsT-transformed black strings from single-trace $T\bar{T}$ at finite temperature. On the other hand, when only one of the $U(1)$ directions belongs to the AdS$_3$ factor, the holographic dual is described by the single-trace $J\bar T$/$T\bar J$ deformation~\cite{Apolo:2019yfj}, a conjecture supported by matching the string theory spectrum at both zero~\cite{Apolo:2018qpq,Chakraborty:2019mdf} and finite temperature~\cite{Apolo:2019yfj}, and reproducing the thermodynamics of the bulk from the boundary side of the correspondence~\cite{Apolo:2019yfj}.\footnote{If both of the TsT directions are taken along translational $U(1)$ isometries of the internal manifold then the deformation of the dual CFT is a marginal $J\bar{J}$ deformation. The spectrum of the deformed theory is partially captured by eqs.~\eqref{tstL02} and~\eqref{tstL0bar2}, with additional equations for the deformed $U(1)$ charges that can be obtained from spectral flow of the symmetry algebra, as discussed in more detail in~\cite{Apolo:2019yfj}.}

We now describe in more detail the proposed duality between TsT transformations of string theory on AdS$_3 \times S^3 \times M^4$ backgrounds and irrelevant deformations of 2d CFTs.


\subsubsection*{$\bm{T\bar{T}}$ deformations}

The single-trace $T\bar{T}$ deformation of the CFT dual to string theory on AdS$_3\times S^3 \times M^4$ is proposed to be the symmetric product $S^p \M_{T\bar T} \equiv (\M_{T\bar T})^p/S_p$ where $\M_{T\bar T}$ is the $T\bar T$ deformation of the seed CFT $\M$~\cite{Giveon:2017nie}. In particular, the action of the deformed seed $\M_{T\bar T}$ is given by eq.~\eqref{cftdeformation} where $\mu \to \mu_{T\bar{T}}$ is a constant with dimensions of length squared and the Noether one-forms  $J^i_{(m)}$ are given in terms of the stress tensor $T^i_{\mu\nu}$ of $\M_{T\bar{T}}$ by
  \eq{
  J^i_{(1)} = T^i_{xx} dx + {T}^i_{\bar{x} x} d\bar{x}, \qquad J^i_{(\bar{2})} = T^i_{x \bar x } dx + T^i_{\bar x \bar x} d\bar{x}. \label{TTbarcurrents}
  }
In eq.~\eqref{TTbarcurrents} $x = \vp + t$ and $\bar{x} = \vp - t$ are the left and right-moving coordinates of the dual field theory, which also correspond to the lightcone coordinates of AdS$_3$. In particular, the Noether currents associated with the $J^i_{(1)}$ and $J^i_{({\bar{2}})}$ one-forms generate translations along the $x$ and $\bar{x}$ coordinates, respectively. The single-trace $T\bar{T}$ deformation is dual to a TsT transformation of string theory on AdS$_3\times S^3 \times M^4$ provided that the TsT coordinates $X^1$ and $X^{\bar{2}}$ are two global $U(1)$ isometries of the locally $AdS_3$ geometry, 
\eq{
\p_{X^1}\in SL(2,R)_L, \qquad \p_{X^{\bar 2}}\in SL(2,R)_R,
}
 where $SL(2,R)_{L/R}$ are the left and right-moving isometries of global AdS$_3$, respectively. This TsT transformation is equivalent to the instantaneous deformation of the worldsheet action given in eq.~\eqref{tstdeformation} where the Noether currents corresponding to the ${\bm j}_{(1)}$ and ${\bm j}_{({\bar{2}})}$ one-forms generate shifts along the target space coordinates ${X}^1$ and ${X}^{\bar{2}}$. As a result, there is a one-to-one correspondence between the $\bm{j}_{(n)}$ and $J_{(n)}^i$ one-forms featured in the worldsheet~\eqref{tstdeformation} and field theory deformations~\eqref{cftdeformation}. Comparison of these equations suggests that the string and field theory deformation parameters $\til{\l}$ and $\mu_{T\bar{T}}$ are related by
  \eq{
  \mu_{T\bar{T}} = \ell^2\til{\l}  \equiv \frac{\ell^2 }{k}\l,  \label{ttbardictionary}
  }
where $\ell$ is the scale of AdS and we have introduced another dimensionless parameter $\l = k\til{\l}$ that is useful in what follows. Note that the AdS scale determines the size of the cylinder at the boundary and that $\ell$ and $\mu_{T\bar{T}}$ are the only dimensionful parameters in the dual field theory.

Evidence for the correspondence between TsT transformations and $T\bar{T}$ deformations follows from matching their spectrum and thermodynamics. The spectrum on the TsT-transformed background is considered in detail in Section~\ref{se:spectrum} while the thermodynamics of $T\bar{T}$ and TsT are derived in Sections~\ref{se:ttbarthermodynamics} and~\ref{se:tstthermodynamics}, respectively. In this section we would like to provide a quick derivation of the spectrum using the general formulae given in eqs.~\eqref{tstL02} and~\eqref{tstL0bar2}. In order to do this it is necessary to choose, in addition to the TsT directions, the coordinates along which strings are allowed to wind. We will consider strings winding along the spatial circle of AdS$_3$, the latter of which are associated to spectrally-flowed representations of $SL(2,R)$. In this case the TsT and winding directions coincide such that the winding parameters featured in eqs.~\eqref{tstL02} and~\eqref{tstL0bar2} satisfy
  \eq{
   w^{(1)} = w^{(\bar{2})} = w.
  }
Let us parametrize the dimensionless momenta along the TsT coordinates in terms of the dimensionful left and right-moving energies $E_L = \frac{1}{2}(E + J/\ell )$ and $E_R = \frac{1}{2} (E - J/\ell)$ where $E$ is the energy and $J$ the angular momentum,  
  \eq{
  p_{(1)} = \ell E_L, \qquad p_{(\bar{2})} = - \ell E_R.
  }
Using eqs.~\eqref{tstL02} and~\eqref{tstL0bar2}, the zero modes of the worldsheet stress tensor are given by
  \eq{
  \hat{L}_0 = &\til{L}_0 +  \ell E_L  w - \frac{2{\l} \ell^2 }{k}E_L E_R, \qquad  \hat{\bar{L}}_0  = \til{\bar{L}}_0 + \ell E_R w - \frac{2\l \ell^2 }{k}E_L E_R. \label{tstspectrum1}
  }
Imposing the Virasoro constraints before and after the deformation allows us to express the deformed left and right-moving energies in terms of the undeformed ones such that
  \eq{
  E_L(0) = E_L(\l) - \frac{2\l \ell }{w k} E_L(\l) E_R(\l), \qquad E_R(0) = E_R(\l) - \frac{2\l \ell}{w k} E_L(\l) E_R(\l). \label{TTbarspectrum1}
  }
Note that this derivation is valid for the TsT transformation of any orbifold of AdS$_3$ that preserves its translational $U(1)$ isometries, including the massless BTZ black hole (Ramond vacuum) as well as generic BTZ black holes.

The spectrum~\eqref{TTbarspectrum1} is consistent with the single-trace $T\bar{T}$ deformation of the symmetric product~\eqref{symprod}. Note that states with $w=-1$ correspond to long strings that wind once around the boundary circle. In the field theory side the spectrum of these states agrees with the untwisted sector of the symmetric product $S^p \M_{T\bar T}$, which corresponds to the spectrum of the $T\bar{T}$-deformed seed theory $\M_{T\bar T}$. Indeed, it is not difficult to check that when $w = -1$, eq.~\eqref{TTbarspectrum1} matches the spectrum of $T\bar{T}$-deformed CFTs on a cylinder of size $2\pi\ell$ and deformation parameter $\mu_{T\bar T} = {\l\ell^2/k}$. On the other hand, the states with $w < -1$ describe strings that wind $|w|$ times along the boundary circle and their spectrum corresponds to the $Z_{|w|}$ twisted sector of $S^p \M_{T\bar T}$~\cite{Giveon:2017myj}.

Interestingly, a different relationship between TsT transformations and $T\bar{T}$ deformations was found in~\cite{Sfondrini:2019smd}. Therein, a TsT-transformed $D+2$ string worldsheet theory was viewed as a non-perturbative $T\bar T$ deformation of a 2d CFT featuring $D$ number of fields and no holographic duality was involved. In contrast, the worldsheet theory studied in this paper describes the bulk theory dual to a $T\bar T$-deformed CFT at the boundary. In other words, we propose that TsT transformations are holographically dual to single-trace $T\bar{T}$ deformations of two-dimensional CFTs. Nevertheless, it would be interesting to explore if there are further connections between the results of this paper and ref.~\cite{Sfondrini:2019smd}.

\subsubsection*{$\bm{J\bar{T}}/\bm{T\bar{J}}$ deformations} 
In analogy with the $T\bar{T}$ deformation, the single-trace $J\bar{T}$ deformation is defined as the symmetric product $({\cal M}_{J\bar T})^p/S_p$ where ${\cal M}_{J\bar T}$ is the $J\bar T$ deformation of the seed CFT $\cal M$. The action of the deformed seed ${\cal M}_{J\bar T}$ is defined by eq.~\eqref{cftdeformation} where $\mu \to \mu_{J\bar{T}}$ is a constant with dimensions of length and the Noether one-forms $J^i_{(n)}$ are given in terms of a $U(1)$ current $J^i_{\mu}$ and the stress tensor $T^i_{\mu\nu}$ of $\M_{J\bar T}$ by  %
  \eq{
  J^i_{(1)} = J^i_x dx + J^i_{\bar{x}} d\bar{x}, \qquad J^i_{({\bar{2}})} = T^i_{x \bar x } dx + T^i_{\bar x \bar x} d\bar{x}. \label{JTbarcurrents}
  }
A similar definition exists for single-trace $T\bar{J}$ deformations and, without loss of generality, we focus only on the $J\bar{T}$ case. In particular, the Noether currents corresponding to the $J^i_{(1)}$ and $J^i_{({\bar{2}})}$ one-forms generate internal $U(1)$ transformations and translations along the $\bar{x}$ coordinate, respectively. As a result, the single-trace $J\bar{T}$ deformation is dual to a TsT transformation where the TsT coordinates $X^1$ and $X^{\bar{2}}$ are identified with $U(1)$ isometries of the internal manifold and the locally AdS$_3$ factor, for example\footnote{Alternatively, we can take one of the TsT coordinates to lie along a $U(1)$ isometry of the $M^4$ manifold. When $M^4 = T^4$ the corresponding $J\bar{T}$ deformation was studied in~\cite{Chakraborty:2018vja}.}
   \eq{
  \p_{{X}^1} &\in SU(2)_L,  \qquad \p_{{X}^{\bar{2}}} \in SL(2,R)_R,
  }
  where $SU(2)_L$ are the left-moving isometries of $S^3$.  
For this choice of TsT coordinates, the Noether one-forms ${\bm j}_{(1)}$ and ${\bm j}_{({\bar{2}})}$ in eq.~\eqref{tstdeformation} are also in one-to-one correspondence with $J^i_{(1)}$ and $J^i_{({\bar{2}})}$ in eq.~\eqref{cftdeformation}. This follows from the fact that the corresponding Noether currents generate $U(1)$ transformations and translations along the $X^{\bar{2}}$ coordinate in the worldsheet and the field theory. The similarities between the worldsheet and field theory deformations suggest once again that the string and field theory parameters $\l$ and $\mu_{J\bar{T}}$ are related by
  \eq{
  \mu_{J\bar{T}} =  \ell \til{\l}= \frac{\ell }{k}\l . \label{jtbardictionary}
  }

The correspondence between TsT transformations and single-trace $J\bar{T}$ deformations was proposed in~\cite{Apolo:2019yfj} after matching the spectrum and thermodynamics in the bulk and boundary theories. In particular, one finds that the string and field theory deformation parameters are indeed related by eq.~\eqref{jtbardictionary}. We conclude this section by showing that the general expressions for the spectrum of strings winding on the TsT transformed background given in eqs.~\eqref{tstL02} and~\eqref{tstL0bar2} reproduce the spectrum of $J\bar{T}$-deformed CFTs. As before, we would like to consider strings that wind along the spatial circle of AdS$_3$, the latter of which is identified with the spatial circle in the dual field theory. This means that there is winding along one of the TsT directions ($X^{{\bar{2}}}$) and an additional $U(1)$ direction denoted by $X^3 \in \textrm{AdS}_3$. The winding parameters featured in eqs.~\eqref{tstL02} and~\eqref{tstL0bar2} satisfy
  \eq{
  w^{(1)} = 0, \qquad w^{({\bar{2}})} = w^{(3)} = w.
  }

An important ingredient in the derivation of the spectrum is an appropriate identification of the deformed and undeformed $U(1)$ charges. The latter were identified in~\cite{Apolo:2019yfj} with the zero modes of the chiral currents generating $U(1)$ transformations before and after the TsT transformation. This choice justifies the following parametrization of the momenta along the $X^1$, $X^{{\bar{2}}}$ and $X^3$ coordinates
  \eq{
  p_{(1)} = \frac{1}{2} \big( q - \l \ell E_R \big), \qquad p_{({\bar{2}})} = -\ell E_R, \qquad p_{(3)} = \ell E_L, \label{jtbarmomenta}
  }
where $q$ denotes the $U(1)$ charge before the deformation.\footnote{In the conventions of~\cite{Apolo:2019yfj}, $X^1$ in this paper is identified with the coordinate $\psi$ of $S^3$ with periodic identification $\psi\sim\psi+4\pi$. In particular, the charge $p_{(1)}$ is associated with the Killing vector $\p_\psi$. This is to be distinguished from the chiral $U(1)$ charge that appears in the definition of the $J\bar{T}$ deformation in~\cite{Guica:2017lia, Chakraborty:2018vja} and which, in the conventions of~\cite{Apolo:2019yfj}, is associated with $2\p_\psi + \lambda \p_v$. The deformed chiral $U(1)$ charge is given by $Q = q - 2\l \ell E_R$ and is related to the momentum $p_{(1)}$ in~\eqref{jtbarmomenta} via $p_{(1)} = \frac{1}{2}(Q + \l \ell E_R)$.} Eqs.~\eqref{tstL02} and~\eqref{tstL0bar2} then yield the following zero modes of the stress tensor in the deformed theory
  \eq{
  \hat{L}_0 = &\til{L}_0 +  \ell E_L  w - \frac{\l \ell }{k} E_R  \big(q - \l \ell E_R  \big), \qquad  \hat{\bar{L}}_0  = \til{\bar{L}}_0 + \ell E_R w - \frac{\l \ell }{k} E_R  \big(q - \l \ell E_R  \big).
  }
Finally, imposing the Virasoro constraints before and after the TsT transformation we find 
  \eq{
  E_L(0) = E_L(\l) - \frac{\l \ell }{wk} E_R  \big(q - \l \ell E_R  \big), \qquad E_R(0) = E_R(\l) - \frac{\l \ell }{wk} E_R  \big(q - \l \ell E_R  \big).  \label{jtbarspectrum}
  }
The worldsheet spectrum~\eqref{jtbarspectrum} is consistent with the spectrum of single-trace $J\bar T$-deformed CFTs on a cylinder of size $2\pi \ell$ and deformation parameter $\mu_{J\bar T} = {\l\ell / k}$. In particular, long strings with $w = -1$ correspond to single particle states in the untwisted sector of the symmetric product $({\cal M}_{J\bar T})^p/S_p$ and their spectrum matches the spectrum of the $J\bar T$-deformed seed theory $\M_{J\bar{T}}$. On the other hand, the worldsheet spectrum~\eqref{jtbarspectrum} with $w <-1$ corresponds to the spectrum of the $Z_{|w|}$ twisted sector of $({\cal M}_{J\bar T})^p/S_p$.


\section{TsT black strings and $T\bar{T}$ deformations} \label{se:tstbtz}
We now consider the TsT transformation of the BTZ black hole that is relevant in the holographic description of $T\bar{T}$-deformed CFTs at finite temperature. In particular, we derive the spectrum of strings winding on the TsT-transformed background and describe in more detail the holographic dictionary relating the worldsheet spectrum to that of single-trace $T\bar{T}$-deformed CFTs. We also show that this background is closely related, by a change of coordinates and a gauge transformation, to another background that can be obtained from a chiral current-current deformation of the worldsheet action.

\subsection{The BTZ black hole}

Let us begin by describing string theory on a BTZ black hole supported by NS-NS flux. We are interested in the six-dimensional part of the AdS$_3 \times S^3 \times M^4$ background such that the metric, $B$-field, and dilaton are given by\footnote{In this section, tilded variables such as $\til{G}_{\mu\nu}$, $\til{B}_{\mu\nu}$, and $\til{\Phi}$ describe the undeformed background fields. However, in contrast to Section~\ref{se:tst}, we do not put tildes on the coordinates to make equations easier to read.}
  \eqsp{
 {d\til{s}^2}{} &=  \ell^2 \bigg \{ {dr^2\over 4 \big (r^2 - 4 T_u^2 T_v^2 \big )} + r du dv + T_u^2 du^2 + T_v^2 dv^2  + d \Omega_3^2 \bigg \}, \\
 \til{B} &= \frac{\ell^2}{4} \big( \cos\t\, d\phi \we d\psi - 2 r du \we dv \big) , \\
  e^{2\til{\Phi}} &= \frac{k}{p}, \label{btzblackstring}
  }
where $k = \ell^2/\ell_s^2$ with $\ell_s$ the string length and $\ell$ the radius of AdS, while $d\Omega_3^2$ is the metric of the 3-sphere that can be written as
  \eq{
  d\Omega_3^2 = \frac{1}{4} \Big [  \( d \psi + \cos\t \, d \phi \)^2 + d\t^2 + \sin^2 \t \, d \phi^2 \Big ].
  }
The lightcone coordinates $u = \vp + t/\ell$ and $v = \vp - t/\ell$ of the AdS$_3$ factor  satisfy
  \eq{
  (u,v) \sim (u + 2 \pi, v + 2\pi),    \label{btzspatialcircle}
  }
which implies that the dual CFT lives in a cylinder of size $2\pi \ell$. On the other hand, the coordinates of the 3-sphere satisfy
  \eq{
   \psi \sim \psi + 4 \pi, \qquad  \t \sim \t + \pi, \qquad \phi \sim \phi + 2 \pi.
   }

The $p$ and $k$ variables parametrizing the dilaton in eq.~\eqref{btzblackstring} correspond to the electric and magnetic charges of the black string (see Appendix~\ref{ap:charges} for our conventions), the latter of which count the number of NS1 and NS5 branes generating the background
  \eq{
  Q_e = p, \qquad Q_m = k.
  }
On the other hand, the $T_u$ and $T_v$ parameters in eq.~\eqref{btzblackstring} are proportional to the dimensionless left and right-moving temperatures of the black string (after normalization by a factor of $1/\pi$). In particular, the background left and right-moving energies can be computed using the covariant formalism reviewed in Appendix~\ref{ap:charges} and are given by
  \eq{
 {\Q}_{L} \equiv  {\Q}_{\p_u} = \frac{c}{6} {T_u^2} \qquad  {\Q}_{R} \equiv  \Q_{-\p_v} = \frac{c}{6}  {T_v^2}, \label{btzcharges}
  }
where $c$ is the central charge of the dual CFT which satisfies
  \eq{
  c = 6 Q_e Q_m = 6 p k.
  }
The metric in eq.~\eqref{btzblackstring} features a horizon at $r_h = 2 T_u T_v$ with entropy
  \eq{
  S_{BTZ} = \frac{\pi c}{3} (T_u + T_v), \label{btzentropy}
  }
and it is not difficult to verify that it satisfies the first law of thermodynamics.

The bosonic part of worldsheet action~\eqref{hataction} reads
  \eq{
  S_0 &=  \frac{k}{2\pi}  \int dz^2 \bigg\{ \frac{\p r \bar{\p} r}{4 \big(r^2 - 4 T_u^2 T_v^2 \big)} + r \p v \bp u + T_u^2 \p u \bar{\p} u   + T_v^2 \p v\bar{\p} v \bigg \} + \til{S}_{\Omega}, \label{btzaction}
  }
where $k$ will be assumed to be large ($k \gg 1$) and $\til{S}_{\Omega}$ denotes the contribution of the 3-sphere. The action~\eqref{btzaction} features Noether currents generating translations along the $u$ and $v$ coordinates, the latter of which correspond to the two $U(1)$ isometries of the AdS$_3$ part of the background. Using the general expression given in eq.~\eqref{noethercurrents} we find
  \eq{
  \til{j}_{\p_u} = 2 T_u \til{j}^1 \bp + k T_u^2 \t_{(u)}, \qquad \til{j}_{\p_v} = 2 T_v \til{\bar{j}}^1 \p - k T_v^2 \t_{(v)}, \label{undefnoether}
  }
where $\til{j}_{\xi}$ denotes the Noether current associated with the Killing vector $\xi$, $\til{j}^1$ and $\til{\bar{j}}^1$ are chiral currents in the WZW formulation of the theory that are  given by\footnote{In the WZW formulation of the theory $\til{j}^1$ and $\til{\bar{j}}^1$ correspond to the spacelike chiral currents
  \eqst{
  \til{j}^1 &= k \tr \big(\p g g^{-1} \tau^1 \big), \qquad \til{\bar{j}}^1 = k \tr \big( g^{-1} \bp g \tau^1 \big),
  }
where the $SL(2,R)$ generators $\tau^a$ can be chosen as $\tau^1 =\frac{1}{2} \s^1$, $\tau^2 = - \frac{i}{2} \s^2$, $\tau^3 = \frac{1}{2} \s^3$ and the $SL(2,R)$ group element satisfies $g = \exp\big(2 T_u u \tau^1\big) \exp \big\{\big[\log \big[ \frac{1}{2} \big( r + \sqrt{r^2 - 4 T_u^2 T_v^2}\,) \big ] - \log(T_u T_v) \big] \tau^3 \big\} \exp \big(2 T_v v \tau^1\big)$.
\label{WZWfootnote}
}
  \eq{
   \til{j}^1(z) = k \bigg(\frac{r \p v}{2T_u} + T_u \p u \bigg), \qquad \til{\bar{j}}^1(\bar{z}) = k \bigg( \frac{r \bp u}{2T_v} + T_v \bp v \bigg),  \label{chiralcurrents}
  }
 and $\t_{(f)}$ is a topological term that satisfies
  \eq{
  \t_{(f)} \equiv \bp f\, \p - \p f\, \bp.
  }
The topological $\t_{(f)}$ terms contribute only total derivative terms to the corresponding Noether charges, i.e.~to the momenta of the string along the $u$ and $v$ coordinates. Nevertheless, these terms are physically meaningful as they contribute in topologically nontrivial sectors of the theory such as the winding string sector.

We have written eq.~\eqref{undefnoether} in a way that makes manifest the fact that the Noether currents $\til{j}_{\p_u}$ and $\til{j}_{\p_v}$ are chiral up to topological terms that are conserved off-shell.  As discussed in Section~\ref{se:tst}, this is a consequence of the fact that the black string background satisfies the constraint~\eqref{Mconstraint} where $m =1 $ and $\bar{m} = \bar{2} $ correspond to the $\til{X}^1 =u$ and $\til{X}^{\bar{2}} = v$ coordinates. In particular, when the temperatures of the BTZ black hole vanish, the Noether currents become chiral. This observation will help us relate TsT transformations to the chiral current-current deformations considered in~\cite{Giveon:2017nie}.


\subsection{The TsT black string} \label{se:tstbackground}

We now perform a TsT transformation along the $\til{X}^1 = u$ and $\til{X}^{\bar{2}} = v$ coordinates of the BTZ black hole by T-dualizing along $u$, shifting $v \to v - \frac{2\l}{k} u$, and T-dualizing along $u$ once more. Note that we have parametrized the TsT transformation by the dimensionless parameter $\l = k \til{\l}$ so that the deformed background fields do not depend explicitly on $k$. The deformed string-frame metric, $B$-field, and dilaton are given by
  \eqsp{
   ds^2 &=  \ell^2 \bigg \{ {dr^2\over 4 \big (r^2 - 4 T_u^2 T_v^2 \big )} + \frac{r du dv+T_u^2 du^2 + T_v^2 dv^2}{1 + 2\l r +4 \l^2 T_u^2 T_v^2} + d \Omega_3^2  \bigg \}, \\
   B &= \frac{\ell^2}{4} \bigg[ \cos\t\, d\phi \we d\psi  - \frac{2(r + 4 \l T_u^2 T_v^2)}{1 + 2\l r + 4\l^2 T_u^2 T_v^2} \,du \we dv \bigg] ,  \label{tstbackground} \\ 
  e^{2\Phi} &= \frac{k}{p}  \big(1 - 4 \l^2 T_u^2 T_v^2 \big) e^{2(\Phi_{\textrm{TsT}} - \phi_0)} = \frac{k}{p}  \bigg( \frac{1 - 4 \l^2 T_u^2 T_v^2}{1 + 2 \l r + 4 \l^2 T_u^2 T_v^2} \bigg) e^{-2\phi_0},
  }
where $\phi_0$ is a constant that may depend on the temperatures and $\Phi_{\textrm{TsT}}$ is the value of the dilaton determined by Buscher's rule~\cite{Buscher:1987sk,Buscher:1987qj}
  \eq{
  e^{-2\Phi_{\textrm{TsT}}} = e^{-2\til{\Phi}} \sqrt{\frac{\textrm{det } \til{G}_{\mu\nu}}{\textrm{det }{G}_{\mu\nu}}} = \frac{p}{k} \big(1 + 2 \l r + 4 \l^2 T_u^2 T_v^2 \big), \label{TsTdilaton}
  }
where $G_{\mu\nu}$ and $\til{G}_{\mu\nu}$ respectively denote the deformed and undeformed string-frame metrics. 

In analogy with the analysis of the warped BTZ black holes dual to $J\bar{T}/T\bar{J}$ deformations at finite temperature~\cite{Apolo:2019yfj}, we assume that the $\vp = \frac{1}{2}(u + v)$ coordinate is compact such that 
  \eq{
  (u,v) \sim (u + 2 \pi, v + 2\pi). \label{tstidentification}
  }
This identification of coordinates guarantees that the dual field theory is defined on the right cylinder for which the spectrum and thermodynamics of the TsT-transformed background match the corresponding quantities in single-trace $T\bar{T}$-deformed CFTs. At large radius, the metric in eq.~\eqref{tstbackground} has vanishing curvature with a linearly decaying dilaton. In particular, due to the identification~\eqref{tstidentification}, the background metric in~\eqref{tstbackground} is asymptotic to $R^{1,1}\times S^1 \times S^3$, instead of $R^{1,2} \times S^3$. For this reason we will refer to the TsT-transformed background as a \emph{TsT black string} instead of a black hole.

It is interesting to note that it is possible to rescale $\l$ away from eq.~\eqref{tstbackground}. This can be achieved by rescaling the coordinates and temperatures such that
  \eq{
  u \to \sqrt{\l} \, u, \qquad v \to \sqrt{\l}  \, v, \qquad r \to \frac{1}{\l}  \,  r, \qquad T_u \to \frac{1}{\sqrt{\l}}  \,  T_u, \qquad T_v \to \frac{1}{\sqrt{\l}}  \, T_v,
  }
while, at the same time, rescaling the size of the spatial circle $\vp \sim \vp + 2\pi \sqrt{\l}$. Alternatively, these equations imply that backgrounds with different values of $\l$ can be interpreted as describing the same solutions with rescaled spatial circles~\eqref{tstidentification}.

The supergravity equations of motion require the dilaton to satisfy Buscher's rule up to a constant, and we have used this freedom to parametrize $\Phi$ in a convenient way by including an additional factor of $(1 - 4 \l^2 T_u^2 T_v^2)e^{-2\phi_0}$. In particular, the dilaton is real provided that
  \eq{
  T_u T_v \le \frac{1}{2\l}, \label{TuTvbound}
  }
which, as we will see in Section~\ref{se:tstthermodynamics}, reproduces the bound on the product of temperatures in $T\bar{T}$-deformed CFTs. Interestingly, when $\phi_0 = 0$ the dilaton in eq.~\eqref{tstbackground} satisfies a similar equation to that dictated by Buscher's rule, except that the deformed metric $G_{\mu\nu}$ is replaced by another closely related background described in Section~\ref{se:tsts}. We also note that when the temperatures and $\phi_0$ vanish, the metric, $B$-field, and dilaton in eq.~\eqref{tstbackground} reduce to the background fields described in ref.~\cite{Giveon:2017nie} which were obtained by a marginal current-current deformation of the worldsheet action (this was also observed in~\cite{Araujo:2018rho}).  

The worldsheet action on the TsT-transformed background can be written as
  \eq{
 S_{\l} & = S_0 + \dd S_{\l}, \label{deformedaction}
 }
where $S_0$ denotes the original action given in eq.~\eqref{btzaction} while $\dd S_\l$ is the contribution from the TsT transformation which reads
  \eq{
  \begin{split}
\dd S_\l &= -\frac{k\l}{\pi}  \int dz^2 \bigg\{ \frac{r + 2\l T_u^2 T_v^2}{1 + 2\l r + 4\l^2 T_u^2 T_v^2}  \big( r \p v \bp u + T_u^2 \p u \bp u + T_v^2 \p v \bp v\big) \\
& \hspace{2.8cm} + \frac{T_u^2 T_v^2 }{1 + 2\l r + 4\l^2 T_u^2 T_v^2}\big(\p u \bp v - \p v \bp u \big) \bigg\} .
\end{split}
  }
After the deformation, the Noether currents generating translations along $u$ and $v$ are respectively given by
  \eq{
  j_{\p_u} &= \frac{1}{1 + 2\l r + 4\l^2 T_u^2 T_v^2} \Big[ 2 T_u \til{j}^1 \bp + k T_u^2 \t_{(u - 2\l T_v^2 v)} \Big], \label{jpu} \\
  j_{\p_v} &= \frac{1}{1 + 2\l r + 4\l^2 T_u^2 T_v^2} \Big[ 2 T_v \til{\bar{j}}^1 \p - k T_v^2 \t_{(v - 2\l T_u^2 u)} \Big], \label{jpv}
  }
where $\til{j}^1$ and $\til{\bar{j}}^1$ were defined in eq.~\eqref{chiralcurrents}. The fate of the chirally conserved currents of the undeformed theory is no longer manifest in these expressions. Nevertheless, the action~\eqref{deformedaction} still features chiral currents but these are not associated with translations along the $u$ and $v$ coordinates. Instead, the chiral currents are associated with $\l$-dependent translations generated by
  \eq{
  \xi_{(u)} = \p_u + 2 \l T_u^2 \p_v, \qquad \xi_{(v)} = \p_v + 2 \l T_v^2 \p_u,
  }
such that the corresponding Noether currents are chiral up to topological terms
  \eq{
  j_{\xi_{(u)}} & = \bigg[ \frac{2 T_u  \til{j}^1 + 2 k \l T_u^2(r \p u + 2 T_v^2 \p v)}{1 + 2 \l r + 4 \l^2 T_u^2 T_v^2}\bigg] \bp + k T_u^2 \t_{(u)},  \label{jxiu} \\
    j_{\xi_{(v)}} &= \bigg[ \frac{2 T_v \til{\bar{j}}^1 + 2 k \l T_v^2(r \bp v + 2 T_u^2 \bp u)}{1 + 2 \l r + 4 \l^2 T_u^2 T_v^2}\bigg] \p - k T_v^2 \t_{(v)}. \label{jxiv}
 }
Finally, we note that the deformed worldsheet action~\eqref{deformedaction} satisfies 
   \eq{
   \frac{\p S_{\l}}{\p \l} =  -\frac{1}{\pi k} \int  {\bm j}_{\p_u} \wedge \bm{j}_{\p_v}, \label{diffdef1}
   }
where ${\bm j}_{\p_u}$ and $\bm{j}_{\p_v}$ are the one-forms associated with the Noether currents $j_{\p_{u}}$ and $j_{\p_{v}}$ via eq.~\eqref{noetherform}. This result is not surprising since we have shown that generic TsT transformations are equivalent to marginal deformations of the worldsheet action by the antisymmetric product of two Noether currents~\eqref{tstdeformation}. In particular, since $j_{\p_u}$ and $j_{\p_v}$ generate shifts along the $u$ and $v$ coordinates --- the lightcone coordinates of the dual field theory --- we see that there is a one-to-one correspondence between the currents in the worldsheet deformation~\eqref{diffdef1} and the currents in the $T\bar{T}$ deformation of the dual field theory~\eqref{cftdeformation}.
 

\subsection{The $T\bar{T}$ spectrum from the worldsheet}  \label{se:spectrum}

We now derive the spectrum of strings winding on the TsT-transformed background~\eqref{tstbackground}. In particular, we describe in more detail the holographic dictionary relating the worldsheet spectrum to that of single-trace $T\bar{T}$-deformed CFTs. Readers satisfied with the general derivation of the spectrum given in Section~\ref{se:tstspectrum} may want to skip this section.

A crucial feature of TsT transformations is that the equations of motion and Virasoro constraints of the deformed theory can be obtained from those of the undeformed theory by a nonlocal change of coordinates~\cite{Frolov:2005dj,Rashkov:2005mi,Alday:2005ww}. For the TsT transformation considered in this section this change of coordinates reads\footnote{Recall that the Noether currents $j_{\p_u}$ and $j_{\p_v}$ of the deformed and undeformed theories are mapped into each other via eq.~\eqref{hatuv} (see eq.~\eqref{currentmap}). Interestingly, the same is true for the chirally conserved currents
  \eqst{
  2 T_u \hat{j}^1  \bp  = j_{\xi_{(u)}} - k T_u^2 \t_{(u)}    \qquad
  2 T_v \hat{\bar{j}}^1 \p =  j_{\xi_{(v)}} + k T_v^2 \t_{(v)} .
  }
However, note that while the chiral currents on the left-hand side are related to translations along $u$ and $v$, the chiral currents on the right-hand side are related to translations along $u + 2\l T_u^2 v$ and $v + 2\l T_v^2 u$.
}
  \eqsp{
  \p \hat{u} &= \p u + \frac{2\l}{k} j_{\p_v}^{\bar{z}}, \qquad  \bp \hat{u}  = \bp u - \frac{2\l}{k}  j_{\p_v}^z, \\ 
  \p \hat{v} &= \p v - \frac{2\l}{k}  j_{\p_u}^{\bar{z}}, \qquad \bp \hat{v} = \bp v + \frac{2\l}{k}  j_{\p_u}^z, \label{hatuv}
  }
where $\hat{u}$ and $\hat{v}$ are the coordinates of the undeformed theory. The change of coordinates~\eqref{hatuv} leads to nonlocal boundary conditions for the $\hat{u}$ and $\hat{v}$ coordinates
  \eq{
  \hat{u} (\s + 2 \pi) = \hat{u}(\s) + 2\pi \g_{(u)}, \qquad \hat{v} (\s + 2 \pi) = \hat{v}(\s) + 2\pi \g_{(v)}, \label{nonlocalbc}
  }
where the $\g_{(u)}$ and $\g_{(v)}$ parameters satisfy
  \eq{
  \g_{(u)} & \equiv \frac{1}{2\pi} \oint \p_\s \hat{u} = w - \frac{2\l \ell E_R}{k},  \qquad \g_{(v)} \equiv \frac{1}{2\pi} \oint \p_\s \hat{v} = w - \frac{2\l \ell E_L}{k}, \label{gamma}
  }
and $E_L$, $E_R$ denote the left and right-moving energies of the deformed theory, the latter of which are defined by
  \eq{
  \ell E_L = \frac{1}{2\pi} \oint \big( j_{\p_u}^z + j_{\p_u}^{\bar{z}} \big), \qquad \ell E_R = -\frac{1}{2\pi} \oint \big( j_{\p_v}^z + j_{\p_v}^{\bar{z}} \big).
  }
In deriving \eqref{gamma} we have assumed that the string winds along the compact $\vp = \frac{1}{2}(u+v)$ coordinate of the deformed background such that  
  \eq{
 \oint \p_\s u = 2 \pi w, \qquad  \oint \p_\s v = 2\pi w.
  }
Classically, $|w|$ counts the number of times the string winds along $\vp$, while quantum-mechanically, $w$ labels the spectrally-flowed representations of $SL(2,R)$ in the undeformed theory. 

The boundary conditions~\eqref{nonlocalbc} induce spectral flow of the underlying $SL(2,R)_L \times SL(2,R)_R$ symmetry algebra of the undeformed theory. In order to see this we note that the twisted boundary conditions~\eqref{nonlocalbc} can be implemented via
  \eq{
  \hat{u} = \til{u} + \g_{(u)} z, \qquad \hat{v} = \til{v} - \g_{(v)} \bar{z}, \label{spectralflowshift}
  }
where $\til{u}$ and $\til{v}$ denote the coordinates of the undeformed theory satisfying trivial boundary conditions. Under spectral flow the chiral WZW currents~\eqref{chiralcurrents} shift according to
  \eq{
  \til{{j}}^1 &\to \hat{{j}}^1 = \til{{j}}^1 + k T_u \g_{(u)}, \qquad \til{\bar{j}}^{1} \to \hat{\bar{j}}^{1} = \til{\bar{j}}^{1} - k T_v \g_{(v)},
  }
such that, in terms of the energies and the winding, their zero modes satisfy
   \eq{
    \frac{1}{2\pi } \oint \til{{j}}^1 & = \frac{1}{2 T_u} \big( \ell E_L +2 \l T_u^2 \ell E_R - k w T_u^2 \big), \\
      \frac{1}{2\pi} \oint \til{\bar{j}}^1 &= -\frac{1}{2 T_v} \big( \ell E_R + 2\l T_v^2 \ell E_L - k w T_v^2 \big).
    }

Although spectral flow does not change the underlying symmetry algebra, it shifts the components of the stress-energy tensor by
  \eq{
  \hat{T} &= \til{T} - 2 T_u \g_{(u)} \til{j}^1 - k T_u^2 \g_{(u)}^{\,2}, \qquad \hat{\bar{T}} = \til{\bar{T}} + 2 T_v \g_{(v)}  \til{\bar{j}}^1 - k T_v^2 \g_{(v)}^{\,2}.
  }
Consequently, we find that the zero modes $\hat{L}_{0} = -\frac{1}{2\pi} \oint \hat{T}$ and $\hat{\bar{L}}_{0} = -\frac{1}{2\pi} \oint \hat{\bar{T}}$ of the stress tensor are given by
  \eq{
  \hat{L}_0 = &\til{L}_0 +  \ell E_L  w - \frac{2\l \ell^2}{k} E_L E_R, \qquad  \hat{\bar{L}}_0  = \til{\bar{L}}_0 + \ell E_R w - \frac{2\l \ell^2}{k} E_L E_R, \label{tstspectrum2}
  }
where $\til{L}_0$ and $\til{\bar{L}}_0$ denote the zero modes before the deformation. The latter satisfy
  \eq{
  \til{L}_0 \Ket{\psi} =  \Big( -\frac{j(j-1)}{k} + N + \dd \Big ) \Ket{\psi}, \qquad \til{\bar{L}}_0 \Ket{\psi}=  \Big (-\frac{\bar{j}(\bar{j}-1)}{k} + \bar{N} + \bar{\dd} \Big )\Ket{\psi}, \label{L0til}
  }
where $j$ is the $SL(2,R)$ weight of the state $\Ket{\psi}$, $(N, \bar{N})$ denote the left and right levels, and $(\dd, \bar{\dd})$ are the contributions from the $S^3 \times M^4$~\cite{Maldacena:2000hw}. Not surprisingly, eq.~\eqref{tstspectrum2} matches the general derivation of the zero modes $\hat{L}_0$ and $\hat{\bar{L}}_0$ given in eq.~\eqref{tstspectrum1}. Imposing the Virasoro constraints before and after the deformation then yields the deformed spectrum reported in eq.~\eqref{TTbarspectrum1}.
  
In our conventions $w < 0$ guarantees that the spectrum of the undeformed theory is bounded from below for both the discrete and principal continuous representations of $SL(2,R)$. We then find that, with the following choice of parameters,
  \eq{
  \l = \ell_s^{-2} \mu, \qquad \ell = R, \label{dictionary}
  }
the worldsheet spectrum~\eqref{tstspectrum2} can be written as
  \eq{
  E(\mu) &= - \frac{|w| R}{2 \mu} \Bigg[ 1 - \sqrt{1 + \frac{4 \mu}{|w| R} E(0) + \frac{4 \mu^2}{w^2 R^4} J(0)^2} \,\Bigg], \qquad J(\mu) = J(0), \label{ttbarspectrum}
  }
where $E = E_L + E_R$ and $J = R(E_L - E_R)$ denote the energy and angular momentum. When $w = -1$, the spectrum~\eqref{ttbarspectrum} matches the spectrum of $T\bar{T}$-deformed CFTs on a cylinder of size $2\pi R$ and deformation parameter $\mu$, which corresponds to the spectrum of single particle states in the untwisted sector of the symmetric product $S^p \M_{T\bar T}$. In particular, note that the identification of the string and field theory deformation parameters in \eqref{dictionary} matches the expression obtained by comparing the instantaneous deformations of the worldsheet and field theory actions~\eqref{ttbardictionary}. Furthermore, we see from eq.~\eqref{ttbarspectrum} that the spectrum of states with $w < - 1$ can be obtained from that with $w = -1$ by rescaling the size of the circle at the boundary such that $R \to |w| R$. The spectrum of these states corresponds to the $Z_{|w|}$ twisted sector of the symmetric product $S^p \M_{T\bar T}$.


\subsection{TsT and current-current deformations} \label{se:tsts}

We now show that the TsT black string considered in the previous section is closely related to another background by a change of coordinates and a gauge transformation. This background can be obtained from a gauged WZW model that we show is equivalent to an exactly marginal current-current deformation of the worldsheet action driven by the product of two chiral currents $\til{j}^1 \til{\bar{j}}^1$. 

Let us start by deriving the background resulting in a current-current ``symmetric" deformation driven by $\til{j}^1 \til{\bar{j}}^1$. Various approaches exist to determine the background fields of such a deformation. A first technique was developed in~\cite{Hassan:1992gi}, where it was conjectured that $O(d,d)$ transformations of the background fields of any WZW model correspond to marginal deformations of the WZW theory by an appropriate combination of left and right moving currents belonging to the Cartan subalgebra. Another way of implementing the symmetric deformation has been put forward in \cite{Giveon:1993ph, Israel:2003ry}, suggesting that the whole deformation line can be realized as an axially and vectorially gauged $\left(SL(2,R) \times U(1)\right)/U(1)$ WZW model, in which the embedding of the dividing group has a component in both factors. We will take the latter approach.

We write a general element  $g \in SL(2,R) \times U(1)$ with coordinates $(u,v,r)$ for the $SL(2,R)$ part (see footnote~\ref{WZWfootnote}) and an additional coordinate $X$ for the $U(1)$ part. The gauged WZW action~\cite{Witten:1991mm} in light-cone gauge can be written as
\begin{equation}\label{GWZW}
 S = S_0 - \frac{k}{4\pi}\int_\Sigma d^2\sigma \,
 \mbox{Tr}(A_-\pp g g^{-1} - \tilde{A}_+ g^{-1} \pmm g + A_- g
 \tilde{A}_+ g^{-1} - \frac{1}{2} A_- A_+ - \frac{1}{2}
 \tilde{A}_- \tilde{A}_+),
 \end{equation}
where $S_0$ is the undeformed WZW action~\eqref{btzaction} while $A_\pm$ and $\tilde{A}_\pm$ are the gauge fields associated with a subgroup $U(1) \subset SL(2,R) \times U(1)$. The latter are expressed as
\begin{equation}
 A_\pm = t_{L} a_\pm , \qquad \tilde{A}_\pm = t_{R}
 {\tilde a}_\pm,
 \end{equation}
 where $t_L$ and $t_R$ are matrix representations of the
 Lie algebra of $U(1)$ embedded in a representation of $SL(2,R) \times U(1)$ while $a_\pm$ and ${\tilde a}_\pm$ depend on the  worldsheet coordinates. 

We perform a so-called axial gauging adapted to the parametrization in footnote~\ref{WZWfootnote}, which corresponds to taking
\renewcommand*\arraystretch{1}
\eq{
t_L = \left(
 \begin{array}{cccc}
0 &&  1 &\hspace{3pt}  0  \\
 1&&  0 &\hspace{3pt}  0  \\
0 && 0 & \hspace{3pt} i \frac{ \sqrt{2} \kappa}{\sqrt{\kappa^2 - 1}}  \\
\end{array}
\right),\qquad t_R = -t_L,
}
where $\kappa$ is a real parameter. The action \eqref{GWZW} is invariant under the action of left and right local symmetries $g \rightarrow e^{\a t_L} g e^{-\a t_R}$, where $\alpha$ is arbitrary function on the worldsheet. Using this local symmetry, we can choose the gauge function $\alpha$ such that $u \, T_u + v \, T_v = 0$. After integrating out the gauge fields, and making the above gauge choice, we can read the background corresponding to the gauged model. For this, we perform a coordinate transformation by setting $X  =  (u' \, T_u + v' \, T_v) \frac{ \sqrt{2} \kappa}{\sqrt{\kappa^2 - 1}}$ and  $u-v=u'-v'$.  The resulting background is that of the symmetric deformation with $\kappa^2 > 1$. 
To reach the line $\kappa^2 < 1$, one would need to perform a vector gauging with $t_R = t_L$ \cite{Detournay:2005fz}. In both cases, we can further redefine 
\eq{
  \l = \frac{1}{2 T_u T_v} \frac{1 - \k}{1 + \k},
  }
such that the background fields read
 \eqsp{
   ds '{}^2&=  \ell^2 \bigg \{  {dr^2\over 4 \big (r^2 - 4 T_u^2 T_v^2 \big )} + \frac{ r (1 + 4 \l^2 T_u^2 T_v^2)du' dv' }{1 + 2\l r +4 \l^2 T_u^2 T_v^2} -  \frac{8 \l T_u^2 T_v^2 du' dv' }{1 + 2\l r +4 \l^2 T_u^2 T_v^2}  + d\Omega^2_3 \\
    & \hspace{1.25cm} + \frac{\big[(1- 2 \l r + 4 \l^2 T_u^2 T_v^2) (T_u^2 du'^2 + T_v^2 dv'^2)}{1 + 2\l r +4 \l^2 T_u^2 T_v^2}  \bigg \}, \\
    \label{tstsbackground} \\
   B' &= \frac{\ell^2}{4} \bigg[ \cos\t\, d\phi \we d\psi  - \frac{2\big[ r (1 + 4 \l^2 T_u^2 T_v^2) + 8 \l T_u^2 T_v^2 \big] }{1 + 2\l r + 4\l^2 T_u^2 T_v^2} \,du' \we dv' \bigg] , \\ 
  e^{2\Phi'} & = \frac{k}{p}  \bigg( \frac{1 - 4 \l^2 T_u^2 T_v^2}{1 + 2 \l r + 4 \l^2 T_u^2 T_v^2} \bigg).
  }
Note that the dilaton is determined by conformal invariance from the one-loop $\beta$-function~\cite{Buscher:1987qj, Witten:1991yr, Kiritsis:1991zt, Giveon:1993ph} and is given by~\cite{Detournay:2005fz}
  \eq{
  e^{2\Phi'} = e^{2\til{\Phi}} \sqrt{\frac{\textrm{det }G_{\mu\nu}'}{\textrm{det }\til{G}_{\mu\nu}}}.
  }

The background~\eqref{tstsbackground} is closely related to the TsT black string~\eqref{tstbackground} considered in the previous section, and it can be obtained from the latter by the change of coordinates
  \eq{
  u' =u - 2\l T_v^2 v, \qquad v'= v - 2\l T_u^2 u, \label{ttbardiff}
  }
followed by the gauge transformation
  \eq{
  B \to B - 2\l T_u^2 T_v^2 du \we dv. \label{ttbargauge}
  }
The value of the dilaton in~\eqref{tstsbackground} matches precisely the one given in eq.~\eqref{tstbackground} with $\phi_0 = 0$, which further justifies our parameterization of the dilaton in~\eqref{tstbackground}. 

The worldsheet action simplifies significantly on the background~\eqref{tstsbackground} and is given by
  \eq{
  S'_\l = S_0 - \frac{k\l}{ \pi} \int d^2 z \bigg[ \frac{\big(r \p v'+ 2T_u^2 \p u'\big) \big(r \bp u' + 2 T_v^2 \bp v' \big)}{1 + 2 \l r + 4 \l^2 T_u^2 T_v^2} \bigg],
  }
where $S_0$ denotes the action before the deformation~\eqref{btzaction} with $u \to u'$ and $v \to v'$. As described earlier, the deformed action $S'_{\l}$ can be interpreted as the result of an instantaneous deformation of the action by the product of two chiral currents
  \eq{
  \frac{\p S'_{\l}}{\p \l} = - \frac{1}{ \pi  k} \int d^2z \, j^1_{\l} \bar{j}^1_{\l}, \label{diffdef2}
    }
where $j^1_{\l}$ and $\bar{j}^1_{\l}$ have been appropriately normalized to satisfy
  \eq{
  j^1_{\l}(z) &= \frac{\sqrt{1 -4 \l^2 T_u^2 T_v^2}}{1 + 2\l r + 4\l^2 T_u^2 T_v^2} \big(2 T_u \til{j}^1\big), \qquad \bar{j}^1_{\l}(\bar{z}) = \frac{\sqrt{1 - 4 \l T_u^2 T_v^2}}{1 + 2\l r + 4\l^2 T_u^2 T_v^2} \big(2 T_v \til{\bar{j}}^1).
  }
In particular, note that the infinitesimal deformation of the action is driven by $\til{j}^1 \til{\bar{j}}^1$, where $\til{j}^1$ and $\til{\bar{j}}^1$ are the chiral currents of the undeformed theory associated with translations along the $u'$ and $v'$ coordinates. Marginal deformations require the presence of $(1,1)$ operators and integrability ensures that the deformed theory, a finite deformation away from the original one, still retains such currents, namely $j_{\l}^1$ and $\bar{j}_{\l}^1$. The latter are no longer associated with translations along the $u$ and $v$ coordinates, however. Instead, up to topological terms, they correspond to Noether currents associated with the following Killing vectors
  \eq{
  \xi'_{(u')} = {\cal N} \bigg( \p_{u'} + \frac{4 \l T_u^2}{1 + 4 \l^2 T_u^2 T_v^2} \p_{v'} \bigg), \qquad \xi'_{(v')} = {\cal N} \bigg( \p_{v'} + \frac{4 \l T_v^2}{1 + 4 \l^2 T_u^2 T_v^2} \p_{u'} \bigg),
  }
where ${\cal N} = (1 + 4 \l^2 T_u^2 T_v^2)/(1 - 4 \l^2 T_u^2 T_v^2)^{3/2}$.

When the temperatures vanish, the TsT black string~\eqref{tstbackground} and the background~\eqref{tstsbackground} coincide. This is compatible with the fact that, in this case, the differential equations for the worldsheet action~\eqref{diffdef1} and~\eqref{diffdef2} agree with one another
  \eq{
 \int {\bm j}_{\p_u} \wedge \bm{j}_{\p_v} =  \int d^2z\, j^1_{\l} \bar{j}^1_{\l} \qquad \textrm{when} \qquad T_u = T_v = 0.
  }
As a result, for the (zero temperature) Ramond vacuum the proposed holographic description of single-trace $T\bar{T}$ deformations is consistent with the analysis of~\cite{Giveon:2017nie}, which considered current-current deformations of the worldsheet action driven by $\til{j}^1 \til{\bar{j}}^1$. A similar observation holds for the holographic interpretation of the $J\bar{T}/T\bar{J}$ deformations, the latter of which reduce to current-current deformations of the worldsheet when the temperatures vanish~\cite{Apolo:2019yfj}.

For generic values of the temperatures, the TsT black string and the background~\eqref{tstsbackground} differ on their global properties. This follows from the fact that the spatial circle of the original BTZ background~\eqref{btzspatialcircle} maps to different circles in the TsT-transformed solution~\eqref{tstbackground} and in~\eqref{tstsbackground}. In particular, the spatial circle determines along which direction the string is allowed to wind, which plays an important role in the derivation of the spectrum. Relatedly, while the gauge transformation~\eqref{ttbargauge} contributes only total derivative terms to the Noether charges, the non-trivial boundary conditions satisfied by winding strings render some of these contributions physical. As a result, the gauge transformation~\eqref{ttbargauge} induces shifts in the momenta of winding strings that affect the expressions for the spectrum. We also note that the choice of spatial circle affects the gravitational charges and the thermodynamics of the TsT-transformed background.

To summarize, the TsT black string~\eqref{tstbackground} and the background~\eqref{tstsbackground} obtained from marginal deformations of the WZW model are locally equivalent. This connection is useful to understand the worldsheet CFT and the integrability of the deformation, perhaps even at the quantum level. At finite temperatures the two backgrounds feature different global properties that affect the winding string spectrum as well as their holographic interpretation. As we will show later, the global properties of~\eqref{tstbackground} play an important role in matching the spectrum and thermodynamics of the black string to the corresponding quantities in single-trace $T\bar{T}$-deformed CFTs. A similar phenomenon is also found in the single-trace $J\bar{T}/T\bar{J}$ deformations studied in~\cite{Apolo:2019yfj}.


\section{Thermodynamics of TsT from $T\bar{T}$} \label{se:thermodynamics}
In this section we study the thermodynamics of single-trace $T\bar T$-deformed CFTs and the TsT-transformed background~\eqref{tstbackground}. First, we derive the entropy and thermal expectation values of the energy and angular momentum in the symmetric product $S^p \M_{T\bar T}$. Previous discussions on the thermodynamics of $S^p \M_{T\bar T}$ can be found in~\cite{Giveon:2017nie}. We then turn to the TsT black string and compute its gravitational charges, entropy, and the thermodynamic potentials compatible with the first law. Finally, we show that the thermodynamics of the TsT black string with fixed electric charge matches the thermodynamics of single-trace $T\bar{T}$-deformed CFTs.

\subsection{Thermodynamics of $T\bar{T}$-deformed CFTs} \label{se:ttbarthermodynamics}

Let us begin by discussing the spectrum of the symmetric product $S^p \M_{T\bar T}$ at large $p$. The symmetric product features states in both twisted and untwisted sectors. In analogy with the spectrum of the undeformed theory~\cite{Argurio:2000tb}, a state in the symmetric product can be viewed as a composite state consisting of $r$ single particle states with $\sum_{i=1}^r n_i = p$ where ${n_i}$ denotes the length of a $Z_{n_i}$ cycle. In the deformed theory, the energies of the $Z_{n_i}$ twisted states with $1 \le n_i \le p$ are given in \eqref{ttbarspectrum} (with $n_i = -w$) and can be written as
  \eq{
  E_{L/R}^{(n_i)}(0) = E_{L/R}^{(n_i)}(\mu) + \frac{2\mu}{n_i R} E_L^{(n_i)}(\mu) E_R^{(n_i)}(\mu). \label{ttbarspectrum3}
  }
In particular, states with $r=p$ and $n_1 = n_2 = \dots = n_p = 1$ belong to the untwisted sector while states with $r=1$ and $n_1 = p$ belong to the maximally-twisted sector of $S^p \M_{T\bar T}$. The total left and right-moving energies of a state in the symmetric product are then given by
  \eq{
  E_L(\mu) = \sum_{i=1}^r E_{L, i}^{(n_i)}(\mu), \qquad E_R(\mu) = \sum_{i=1}^r E_{R, i}^{(n_i)}(\mu). \label{totalenergy}
  }

The deformed energies of the ground state are obtained by setting each $n_i$ in eq.~\eqref{totalenergy} to $n_i = 1$ and setting the undeformed energies in eq.~\eqref{ttbarspectrum3} to their ground state values $R E^{(1), vac}_L (0) = R E^{(1), vac}_R (0) = -{c_\M}/24=-k/4$ such that
  \eq{
  E^{vac}_L (\mu) = E^{vac}_R (\mu) = -\frac{p R}{4\mu} \bigg(1-  \sqrt{1 - \frac{2 \mu k}{R^2}}\, \bigg). \label{ttbarvacuum}
  }
Note that the total energies of the vacuum before the deformation are given by  $R E^{vac}_L (0) = R E^{vac}_R (0) = -{p k}/4 = -c /24$ as expected. When $\mu>0$, the spectrum of the deformed theory is well behaved provided that 
\eq{
\mu \le \mu_c \equiv \frac{R^2}{2k}.
}
Otherwise, the energy of the ground state, and possibly an additional number of excited states, becomes complex. At the critical value $\mu = \mu_c$, the deformed ground state energies are real and given by $RE^{vac}_L(\mu_c) =R E^{vac}_R (\mu_c) = -c/12$. Furthermore, using the holographic dictionary \eqref{dictionary}, we find that the critical value of the deformation parameter in the string theory formulation of the $T\bar{T}$ deformation is given by
  \eq{
  \l_c = \frac{1}{2}, \label{lambdacritical}
  }
such that $\l \le \l_c$ leads to a well-defined worldsheet spectrum.

Let us now describe the asymptotic density of states in each of the $Z_{n_i}$ twisted sectors of $S^p \M_{T\bar T}$.  Before the deformation, the energies of $Z_{n_i}$ twisted states are related to the energies of untwisted states by $E^{(n_i)}_{L/R}(0) = (1/n_i) {E^{(1)}_{L/R}(0)}$. In particular, the energy gap in the twisted sector is rescaled by a factor of $1/n_i$, leading to the modified Cardy regime  
\eq{
RE_{L/R}^{(n_i)}(0) \gg \frac{k}{n_i}. \label{entropyregime}
}
Clearly, the  maximally-twisted sector falls within the Cardy regime at the lowest energy, a property that is preserved after the deformation as shown in detail later (see eq.~\eqref{deformedregime}). Moreover, from the spectrum in \eqref{ttbarspectrum3} we see that there is no level crossing between states in the symmetric product $S^p \M_{T\bar T}$, implying that the density of states does not change after the $T\bar{T}$ deformation. As a result, the logarithm of the density of $Z_{n_i}$ twisted states is given by
\eq{
\!\!S^{(n_i)}_{T\bar{T}} &= 2\pi \bigg[ \sqrt{\frac{n_i c_{\M}}{6} R E_L^{(n_i)}(0)} + \sqrt{\frac{n_i c_{\M}}{6} R E_R^{(n_i)}(0)} \,\bigg],  \notag \\
& = 2\pi \bigg [\sqrt{n_i k R E_L^{(n_i)}(\mu) \Big[1 \!+\! \frac{2\mu}{n_i R} E_R^{(n_i)}(\mu) \Big]} + \sqrt{n_i k R E_R^{(n_i)}(\mu) \Big[ 1 \!+\! \frac{2\mu}{n_i R} E_L^{(n_i)}(\mu) \Big]}\, \bigg ],  \label{entropytwisted2}
}
which is the Cardy formula written in terms of the deformed spectrum \eqref{ttbarspectrum3}.

Given a choice of $\{n_i\}$, the degeneracy of states in $S^p \M_{T\bar T}$ can be obtained by summing over the contributions from each of the $Z_{n_i}$ twisted sectors such that
   \eq{
   S^{\{n_i\}}_{T\bar{T}}(E_L, E_R) &= \sum_{i=1}^r S^{(n_i)}_{T\bar{T}} \big(E^{(n_i)}_L, E^{(n_i)}_R \big). \label{totalentropy}
  }
In the microcanonical ensemble the total left and right-moving energies \eqref{totalenergy} are held fixed and assumed to be large.  We then note that the following partition of the energies among the different twisted sectors
  \eq{
  E^{(n_i)}_{L/R}(\mu) = \frac{n_i}{p} E_{L/R}(\mu),   \label{energypartition}
  }
extremizes the total entropy in eq.~\eqref{totalentropy}. Furthermore, it is not difficult to show that the entropy~\eqref{totalentropy} satisfies $(\p_{E_L}^2 S)(\p_{E_R}^2 S) > (\p_{E_L} \p_{E_R} S)^2$ and $\p_{E_L}^2 S < 0$. As a result, the partition of the energies  in \eqref{energypartition} is a local maximum of the entropy. Using eq.~\eqref{energypartition} and Cardy's formula \eqref{entropytwisted2} in each twisted sector, we find that the total entropy \eqref{totalentropy}  does not depend on the choice of  $\{n_i\}$ and is given by
  \eq{
  S_{T\bar{T}}(E_L, E_R) & = 2\pi \bigg [ \sqrt{\frac{c}{6} R E_L(\mu) \Big[1 + \frac{2\mu}{R p} E_R(\mu) \Big]} + \sqrt{\frac{c}{6} R E_R(\mu) \Big[ 1 + \frac{2\mu}{Rp} E_L(\mu) \Big]}\, \bigg ]. \label{entropysymprod}
 }

It is interesting to note that the distribution of energies in \eqref{energypartition} that maximizes the entropy can be translated, by means of eq.~\eqref{ttbarspectrum3}, into a similar distribution for the undeformed states, namely $E_{L/R}^{(n_i)}(0)= (n_i/p) E_{L/R}(0)$ where $E_{L/R}(0) = \sum_{i=1}^r E_{L/R}^{(n_i)}(0)$ are the total energies in the undeformed theory. Consequently, we find that the total energies of the symmetric product before and after the deformation are related by 
\eq{
 E_{L/R}(0) = E_{L/R} (\mu) + \frac{2\mu}{p R} E_L(\mu) E_R(\mu), \label{ttbarspectrum4}
 } 
and the regime \eqref{entropyregime} can be written as%
\eq{ 
RE_{L/R}(0) \gg  \frac{c}{n_i^2}>{c\over p^2}. \label{entropyregime2} 
} 
The entropy formula \eqref{entropysymprod} is valid as long as the energies satisfy eq.~\eqref{entropyregime2}.\footnote{It may be possible to enlarge the range of validity of \eqref{entropysymprod} using the arguments of \cite{Hartman:2014oaa}.} Note that when $p$ is large, the maximally twisted sector dominates the entropy at high, but not very high energies. This occurs, for example, when $RE_{L/R}(0) = (c/6)[\pi RT_{L/R}(0)]^2 \gg {c / p^2}$ where $T_{L/R}(0)$ are the undeformed temperatures and $RT_{L/R}(0)$  are parametrically of order one. As we increase the temperatures, more choices of $\{n_i\}$ contribute to the entropy. In particular, $RT_{L/R}(0)\gg 1$ corresponds to the usual Cardy regime $RE_{L/R}(0)\gg c$ where all possible choices of $\{n_i\}$ contribute to the entropy.

Let us now express the regime of validity \eqref{entropyregime2} of the entropy in terms of the deformed energies. The deformed spectrum is a monotonic function of the undeformed energies, which means that we only need to express the threshold value ${c/p^2}$ on the right hand side of \eqref{entropyregime2} in terms of the deformed spectrum. Using eq.~\eqref{ttbarspectrum4}, we find that the deformed threshold remains ${c/p^2}$ up to corrections of order $\mathcal{O}({\mu k/ p^2})$. As a result, the entropy \eqref{entropysymprod} is valid provided that deformed energies satisfy 
\eq{
R E_{L/R}(\mu) \gg {c\over p^2}. \label{deformedregime}
}
Interestingly, when the undeformed energy satisfies $c/\mu k \gg RE(0)\gg c/p^2$, the deformed energy falls within the same regime $c/\mu k \gg RE(\mu)\gg c/p^2$ (where we have set the angular momentum to zero for convenience). In this range, the entropy \eqref{entropysymprod} is in the Cardy regime where $S \sim 2\pi \sqrt{\frac{c}{3} R E(\mu)}$. When the undeformed energy $RE(0) $ is larger than or comparable to ${c /\mu k}$, the deformed energy $R E (\mu)$ is also larger than or comparable to ${c /\mu k}$. In this case, the entropy \eqref{entropysymprod} falls within the Hagedorn regime where $S \sim 2\pi \sqrt{2 \mu k} E(\mu)$.

We can also write down the entropy of the symmetric product $S^p \M_{T\bar T}$ in the canonical ensemble. Using the first law of thermodynamics, the temperatures $T_L \equiv \big(\p S_{T\bar T}/\p E_L\big)^{-1}$ and $T_R \equiv \big(\p S_{T\bar T}/\p E_R \big)^{-1}$ conjugate to the left and right-moving energies are given by
   \eq{
    \frac{1 }{\pi T_L(\mu)} &= \sqrt{\frac{k \big[p R +  2 \mu  E_R(\mu)\big]}{E_L(\mu)}}+ 2\mu k\sqrt{\frac{E_R(\mu)}{k\big[p R + 2 \mu E_L(\mu)\big]}}, \label{ttbartemperatures1}\\
   \frac{1 }{\pi T_R(\mu)} &= \sqrt{\frac{k \big[p R +  2 \mu  E_L(\mu)\big]}{E_R(\mu)}}+ 2\mu k\sqrt{\frac{E_L(\mu)}{k\big[p R + 2 \mu E_R(\mu)\big] }}.\label{ttbartemperatures2}
  } 
 The energies on the right-hand side of eqs.~\eqref{ttbartemperatures1} and~\eqref{ttbartemperatures2} should be understood as thermal expectation values in the deformed theory at temperatures $T_{L}(\mu)$ and $T_{R}(\mu)$. Furthermore, these equations imply that, in the high $E_{L}$ and $E_{R}$ limits, the product of the left and right-moving temperatures is bounded from above by
  \eq{
  T_L(\mu) T_R(\mu)  \le  \frac{1}{8\pi^2 \mu k},   \label{ttbarbound}
  }
  where the bound is reached as $E_{L/R}(\mu)\to \infty$. Finally, using eqs.~\eqref{ttbartemperatures1} and~\eqref{ttbartemperatures2} we find that in the canonical ensemble the entropy is given by
 \eq{
  S_{T\bar{T}}(T_L, T_R) = \frac{\pi^2 R c}{3} \Bigg[ \frac{T_L(\mu) + T_R(\mu)}{\g} \Bigg], \label{entropysymprod2}
  }
where $\g$ is defined by
  \eq{
  \g \equiv \sqrt{1 - (8\pi^2 \mu k)T_L(\mu) T_R(\mu)}. \label{gammaparameter}
  }
In particular, note that the bound on the product of the temperatures~\eqref{ttbarbound} implies that $0 \le \gamma\le1$ such that the entropy is real when $\mu \ge 0$. In analogy with the discussion in the microcanonical ensemble, the density of states exhibits Cardy growth if $ 1/\mu k \gg T_L(\mu) T_R(\mu) \gg {1 /R^2 p^2}$ and Hagedorn growth if $T_L(\mu) T_R(\mu)$ is comparable to $1/\mu k$.


\subsection{Thermodynamics of TsT black strings} \label{se:tstthermodynamics}

The TsT black string is a solution to the supergravity equations of motion for any constant values of $T_u$, $T_v$, $\phi_0$, and the deformation parameter $\l$. We are interested in computing the gravitational charges on the space of solutions parametrized by these variables. Using the covariant formulation of charges reviewed in Appendix~\ref{ap:charges}, we find that the left and right-moving energies associated with the Killing vectors $\p_u$ and $-\p_v$ are integrable only for fixed values of $\l$. The finite expressions for these charges are given by
  \eq{
  \Q_L \equiv \Q_{\p_u} = \frac{c}{6} \frac{ \big ( 1 + 2\l T_v^2\big) T_u^2}{1 - 4\l^2 T_u^2 T_v^2} e^{2\phi_0} , \qquad \Q_R \equiv \Q_{-\p_v} = \frac{c}{6} \frac{ \big(1 + 2\l T_u^2\big) T_v^2}{1 - 4\l^2 T_u^2 T_v^2} e^{2\phi_0},\label{charges}
  }
where the factor of $e^{2\phi_0} /(1 - 4\l^2 T_u^2 T_v^2)$ originates from the shift of the dilaton accompanying the $e^{2 \Phi_{\textrm{TsT}}}$ term in eq.~\eqref{tstbackground}. Note that the bound~\eqref{TuTvbound} guarantees that the energies of the black string do not become negative as we increase the temperatures. In Section~\ref{se:ttbarflow} we will consider a different parametrization of the black string where the charges are manifestly well-behaved but the TsT origin of the spacetime is obscured.  

The background~\eqref{tstbackground} also features electric and magnetic charges that are given by
  \eq{
  Q_e = p e^{2\phi_0}, \qquad Q_m = k. \label{electriccharge}
  }
We recall that before the TsT transformation, the BTZ black hole can be obtained from the near-horizon limit of $k$ coincident NS5 branes and $p$ coincident strings such that $Q_e = p$ and $Q_m = k$. Thus, the values of the electric and magnetic charges are preserved after the TsT deformation provided that $\phi_0 = 0$. This is one of the motivations behind our parametrization of the dilaton in eq.~\eqref{tstbackground}.  
  
Let us now turn to the thermodynamics of the TsT black string. The background~\eqref{tstbackground} features a horizon at $r_h = 2 T_u T_v$ whose near-horizon ($r - r_h \equiv 4 T_u T_v \e^2 \to 0$) geometry is described by
  \eq{
  \!\! ds^2 =  \ell^2 e^{\phi_0} \sqrt{\frac{p}{k}\bigg( \frac{1 + 2 \l T_u T_v}{1 - 2 \l T_u T_v} \bigg)} \Bigg[ d\e^2 - \e^2 \bigg(\frac{T_u du - T_v dv}{1 + 2\l T_u T_v} \bigg)^2 \! \!+\! \bigg(\frac{T_u du + T_v dv}{1 + 2\l T_u T_v}\bigg)^2 \Bigg] \!+ \dots \label{nearhorizon}
  }
Regularity of the Euclidean continuation of eq.~\eqref{nearhorizon} implies that the $u$ and $v$ coordinates satisfy the thermal identification
  \eq{
  (u,v) \sim (u + i/\ell T_L, v - i/\ell T_R),
  }
where the physical left and right-moving temperatures $T_L$ and $T_R$ satisfy
  \eq{
  \ell T_L = \frac{1}{\pi} \frac{T_u}{1 + 2 \l T_u T_v}, \qquad  \ell T_R = \frac{1}{\pi} \frac{T_v}{1 + 2 \l T_u T_v}. \label{temperatures}
  }
From the near-horizon metric~\eqref{nearhorizon} we readily find that the entropy of the TsT black string in terms of the $T_u$ and $T_v$ parameters is given by
  \eq{
  S_{\textrm{TsT}}(T_u, T_v, Q_e)= 2\pi Q_e Q_m  \bigg( \frac{T_u + T_v}{1 - 2 \l T_u T_v} \bigg). \label{tstentropy}
  }
Alternatively, we can write the entropy of the black string entirely in terms of the conserved charges such that
\eq{
S_{\textrm{TsT}}(\Q_L, \Q_R, Q_e) & = 2\pi \Big \{ \sqrt{   \Q_{L} \big ( Q_e Q_m + 2\l  \Q_{R} \big)} +\sqrt{   \Q_{R} \big ( Q_e Q_m + 2\l  \Q_{L} \big)}\, \Big \}. \label{tstentropy2}
}
We interpret eq.~\eqref{tstentropy2} as the entropy of the black string in the microcanonical ensemble.

When neither the $T_u$ or $T_v$ parameters are zero, the energies~\eqref{charges} and entropy~\eqref{tstentropy} of the black string are positive thanks to the bound $T_u T_v \le 1/2\l$ given in eq.~\eqref{TuTvbound}, where the equal sign corresponds to the case where both the energies and the entropy become infinite. This bounds leads to the following bound on the product of the physical temperatures
  \eq{
  T_L T_R \le \frac{1}{8\pi^2\ell^2 \l}, \label{tstbound}
  }
which, using the holographic dictionary~\eqref{dictionary}, is the same bound~\eqref{ttbarbound} obtained from the field theory analysis of $T\bar{T}$-deformed CFTs at finite temperature. Hence, the parametrization of the black string in terms of the $T_u$ and $T_v$ variables is compatible with the fact that $T\bar{T}$-deformed CFTs feature a maximum product of temperatures.

If the electric charge is allowed to vary, i.e.~if $\d \phi_0 \ne 0$, then the black string features a chemical potential conjugate to $Q_e$ that is given by~\cite{Compere:2007vx}
  \eq{
  \nu = \pi k (T_u + T_v).
  }
It is then not difficult to show that the thermodynamic potentials and their conjugate charges satisfy the first law of thermodynamics,
  \eq{
  \d S_{\textrm{TsT}}(\Q_L, \Q_R, Q_e) = \frac{1}{\ell T_L} \d \Q_{L} + \frac{1}{\ell T_R} \d\Q_{R} + \nu \d Q_e.
  }
Finally, we note that it is possible write the physical charges~\eqref{charges} in terms of the physical temperatures~\eqref{temperatures} without using the intermediate $T_u$ and $T_v$ variables. However, it turns out to be more convenient to express the temperatures in terms of the charges, the latter of which can be shown to satisfy
  \eq{
  \frac{1}{\pi \ell T_L} &= \sqrt{\frac{Q_e Q_m + 2 \l \Q_R}{\vphantom{Q_e Q_m}\Q_L}} + 2\l \sqrt{\frac{\Q_R}{Q_e Q_m + 2\l \Q_L}}, \label{tsttemperatures1} \\
    \frac{1}{\pi \ell T_R} &= \sqrt{\frac{Q_e Q_m + 2 \l \Q_L}{\vphantom{Q_e Q_m}\Q_R}} + 2\l \sqrt{\frac{\Q_L}{Q_e Q_m + 2\l \Q_R}}. \label{tsttemperatures2}
  }
  %


\subsection{Matching the thermodynamics} \label{se:tstmatching}

We have seen that the TsT black string is characterized by three independent charges that correspond to the energy, angular momentum, and the electric charge. The black string also features conjugate thermodynamic potentials which suggest that this background is dual to a thermal state in the dual field theory. On the other hand, the thermodynamics of $T\bar{T}$-deformed CFTs were derived in the absence of any additional thermodynamic potentials other than the temperature and the angular potential. For this reason, we restrict the analysis of this section to fixed values of the electric charge such that $\d \phi_0 = 0$. Furthermore, without loss of generality, we assume that $\phi_0 = 0$, which implies that the values of the electric and magnetic charges are unchanged by the deformation and satisfy $Q_e = p$ and $Q_m = k$. As a result, the central charge of the CFT dual to string theory on AdS$_3 \times S^3 \times M^4$ is preserved after the TsT transformation such that $c = 6 Q_e Q_m = 6 p k$.\footnote{For any other values of $\phi_0$ satisfying $\d \phi_0 = 0$, the entropy of the TsT black string matches the entropy of single-trace $T\bar{T}$-deformed CFTs with a rescaled value of the central charge given by $c_{\phi} = 6 Q_e Q_m = 6 p k e^{2\phi_0}$.}

Let us first consider the microcanonical ensemble. In this ensemble we identify the thermal expectation values of the dual $T\bar{T}$-deformed CFT with the energies of the TsT black string, namely
   \eq{
      R E_L &= {\Q}_{L}, \qquad R E_R = {\Q}_{R}. \label{bulkboundarycharges}
  }
The canonical and microcanonical expressions for the entropy derived in Section~\ref{se:ttbarthermodynamics} are valid in the regime~\eqref{deformedregime} where the energies are assumed to be large. In this limit, we find that the entropy of single-trace $T\bar{T}$-deformed CFTs in the microcanonical ensemble~\eqref{entropysymprod} matches the Bekenstein-Hawking entropy of the black string given in \eqref{tstentropy2},
	\eq{
	S_{T\bar T} (E_L, E_R) =  S_{\textrm{TsT}} (\Q_{L}, \Q_{R}, p),
	}
where we used eq.~\eqref{bulkboundarycharges} and the holographic dictionary~\eqref{dictionary} relating the bulk and boundary deformation parameters.

We now turn to the canonical ensemble. We expect the TsT black string to correspond to a thermal state in the dual field theory whose physical left and right-moving temperatures $T_L$ and $T_R$ are determined from the horizon generator by eq.~\eqref{temperatures}. Interestingly, the structure of $T_L$ and $T_R$ is a special one as it guarantees that the $\g$ parameter introduced in the field theory analysis in eq.~\eqref{gammaparameter} is a rational function, namely
  \eq{
  \g = \sqrt{1 - 8\pi^2 \mu k T_L T_R } = \frac{1 - 2 \l T_u T_v}{1 + 2 \l T_u T_v}. 
  }
We then find that the entropy of single-trace $T\bar{T}$-deformed CFTs at temperatures~\eqref{temperatures} also matches the entropy of TsT black strings in eq.~\eqref{tstentropy}, 
  \eq{
  S_{T\bar{T}}(T_L, T_R) = S_{\textrm{TsT}}(T_u, T_v, p), 
  }
where once again we used the dictionary~\eqref{dictionary} relating the string and field theory variables. 

The fact that we are able to match the entropy in both the canonical and microcanonical ensembles suggests that the physical temperatures and the gravitational charges of the TsT black string are related via the thermodynamic relations of single-trace $T\bar{T}$-deformed CFTs given in eqs.~\eqref{ttbartemperatures1} and~\eqref{ttbartemperatures2}. Indeed, it is not difficult to verify using eq.~\eqref{bulkboundarycharges} that the physical temperatures and charges of the black string related via eqs.~\eqref{tsttemperatures1} and~\eqref{tsttemperatures2} reproduce the corresponding equations in $T\bar{T}$-deformed CFTs. This provides further evidence that the matching of the entropy in the canonical and microcanonical ensembles is consistent. Thus, we conclude that the TsT black string with fixed electric charge is dual to a thermal state in a single-trace $T\bar{T}$-deformed CFT.
 

\section{The phase space} \label{se:phasespace}

In this section we take a closer look at the phase space of solutions obtained from a TsT transformation of the BTZ black hole. To begin, we find an everywhere-smooth background that interpolates between the global AdS$_3$ vacuum in the IR and a linear dilaton background in the UV. This solution is obtained by analytic continuation of the $T_u$ and $T_v$ parameters and corresponds to the NS-NS vacuum of the dual $T\bar{T}$-deformed CFT. We then show that, in order to recover the $T\bar{T}$ flow between states in the space of BTZ and TsT black strings solutions, it is necessary to consider $\l$-dependent $T_u$ and $T_v$ parameters. Finally, we comment on the solutions with $\l <0$, which corresponds to the superluminal sign of the $T\bar{T}$ deformation in the field theory.

\subsection{The ground state} \label{se:NSvacuum}
In previous sections we had implicitly assumed that the $T_u$ and $T_v$ parameters are real so that the corresponding black string solutions describe thermal states in the dual $T\bar{T}$-deformed CFT. In particular, the background with $T_u = T_v = 0$ features vanishing energies and describes the R-R vacuum in the dual field theory. A natural question to ask is what is the holographic dual to the NS-NS vacuum state characterized by left and right-moving energies given in eq.~\eqref{ttbarvacuum}. In this section we find the NS-NS ground state and rederive the critical value $\l_c$~\eqref{lambdacritical} that guarantees that the vacuum energy is real.

Let us begin by performing an analytic continuation of the temperatures
  \eq{
  T_u^2 = T_v^2 \equiv -\rho_0^2, \qquad  \rho_0^2 > 0, \label{ancont}
  }
after which the TsT-transformed background in eq.~\eqref{tstbackground} no longer features a horizon or angular momentum. The corresponding background features a conical defect/surplus at $r = 2 \rho_0^2$ unless the following condition is satisfied
  \eq{ 
  {4\rho_0^2\over (1+2\lambda \rho_0^2)^2} = 1. \label{rhoconstraint}
  }
The solution to eq.~\eqref{rhoconstraint} that guarantees smoothness of the TsT-transformed spacetime is given by
  \eq{
  \rho_0^2={1-\lambda - \sqrt{1-2\lambda } \over 2\lambda^2} = \bigg[{1\over 2\lambda} \big(1-\sqrt{1-2\lambda}\,\big) \bigg]^2,  \label{rho0}
  }
where we have chosen the branch satisfying $\rho_o^2 \to 1/4$ as $\l \to 0$, which corresponds to the value of $\rho_0$ for global AdS$_3$ in this gauge. 

Eq.~\eqref{rho0} admits real solutions only when $\l \le \l_c = 1/2$. This is the same critical value of the deformation parameter that guarantees a real  ground state energy in the dual $T\bar{T}$-deformed CFT, cf.~eq.~\eqref{lambdacritical}. This is not a coincidence, as the left and right-moving energies of the TsT-transformed background satisfying eqs.~\eqref{ancont} and~\eqref{rho0} match the energies of the vacuum state in single-trace $T\bar{T}$-deformed CFTs. In order to see this, we note that the background left and right-moving energies satisfy 
   \eq{
   \Q^{vac}_L = \Q^{vac}_R = -\frac{c}{6} {\rho_o^2\over1+2\lambda \rho_o^2} = - {c\over 24\lambda} \big(1-\sqrt{1-2\lambda}\,\big), \label{chargesNSvacuum}
   }
  where, following the discussion in Section~\ref{se:tstmatching}, we have set $\phi_0 = 0$. Using the holographic dictionary~\eqref{dictionary} we find that these expressions match the vacuum energies of single-trace $T\bar{T}$-deformed CFTs given in eq.~\eqref{ttbarvacuum}. Thus, the TsT background satisfying eqs.~\eqref{ancont} and~\eqref{rho0} is dual to the NS-NS ground state in the dual $T\bar{T}$-deformed theory.

Using the following change of coordinates
  \eq{
  r = 2 \rho_o^2 \big(2\rho^2 + 1),
  }
the NS-NS vacuum can be written as
 \eqsp{
   ds^2 &=  \ell^2 \bigg \{ {d\rho^2\over \rho^2 +1  } + \frac{\rho^2 d\varphi^2 -(\rho^2+1)dt^2 }{1+ 2\l \rho^2} + d \Omega_3^2  \bigg \}, \\
   B &= \frac{\ell^2}{4} \bigg[ \cos\t\, d\phi \we d\psi  - \frac{2( \rho^2 + \rho_o)}{1+2\l \rho^2} \,du \we dv \bigg] , \label{NSvacuum} \\
  e^{2\Phi} &=  \frac{k}{p} \frac{(1 - 2 \l \rho_o^2) }{2\rho_o(1+ 2\l \rho^2)}e^{-2\phi_0}.
  }
In this gauge, it is clear that the string frame metric goes to $R^{1,1} \times S^1 \times S^3$ as $\rho \to \infty$, while the geometry caps off at $\rho = 0$ where it approaches global AdS$_3 \times S^3$ in the region $\lambda \rho^2 \ll1$. Consequently, the background~\eqref{NSvacuum} is a smooth solution to the supergravity equations of motion interpolating between global AdS$_3$ in the IR and a linear dilaton in the UV. We also note that the solution~\eqref{NSvacuum} depends on $\rho_0$ only through the $B$ field and a constant shift in the dilaton. In particular, when $\l \le \l_c$ all of the background fields are real. Interestingly, when $\l > \l_c$ eq.~\eqref{NSvacuum} remains a solution to the equations of motion but the $B$ field and dilaton become complex.  Finally, we note that at the critical value $\l = \l_c$ we have $\rho_0 = 1$ and the string coupling is zero everywhere. In this special case the supergravity approximation breaks down even in the IR (AdS$_3$) region and we need the Little String Theory description everywhere.

\subsection{$T\bar{T}$ flow from TsT} \label{se:ttbarflow}
 
We have seen that the entropy and thermodynamics of the TsT black string match those of single-trace $T\bar{T}$-deformed CFTs at finite temperature. As discussed in Section~\ref{se:ttbarthermodynamics}, the density of states is not expected to change after the deformation provided that the energies are related by eq.~\eqref{ttbarspectrum3}. An interesting question to ask is how the deformed $T\bar{T}$ spectrum arises in the bulk. A naive guess is to compare the energies before and after the TsT transformation with fixed values of $T_u$ and $T_v$. Given the parametrization of the deformed energies in eq.~\eqref{charges}, the values of the undeformed charges required to satisfy the $T\bar{T}$ spectrum are $\l$-dependent and given by
  \eq{
  \Q_{L}(0) = \frac{c}{6} \bigg[ \frac{(1 + 2 \l T_v^2) T_u}{1 - 4 \l^2 T_u^2 T_v^2} \bigg]^2, \qquad   \Q_{R}(0)  = \frac{c}{6} \bigg[ \frac{(1 + 2 \l T_u^2) T_v}{1 - 4 \l^2 T_u^2 T_v^2} \bigg]^2,
  }
where, following the discussion in Section~\ref{se:tstmatching}, we have set $\phi_0 = 0$. The equations above differ from the charges of the original BTZ black hole before the TsT transformation, see eq.~\eqref{btzcharges}. In other words, while the TsT transformation maps the phase space of the undeformed theory to the phase space of  a $T\bar{T}$-deformed CFT, the flow induced by $\lambda$ with fixed ($\l$-independent) values of $T_u$ and $T_v$  does not agree with the $T\bar{T}$ flow of states.
 
In order to see the $T\bar{T}$ flow, the $T_u$ and $T_v$ variables parametrizing the black string must depend on $\l$ such that
    \eq{
  T_u^2 = \frac{\cE_L(\l)}{1 + 2 \l \cE_R(\l)}, \qquad T_v^2 = \frac{\cE_R(\l)}{1 + 2 \l \cE_L(\l)}, \label{inverseTuTv3}
  }
where the $\cE_L(\l)$ and $\cE_R(\l)$ variables implement the $T\bar T$ flow via 
 \eq{
  \cE_L(0) = \cE_L(\l)\big[1 + 2\l \cE_R(\l)\big], \qquad \cE_R(0) = \cE_R(\l) \big [1 + 2 \l \cE_L(\l)\big].\label{floweq}
  }
The TsT black string~\eqref{tstbackground} is written in terms of $\cE_L(\l)$ and $\cE_R(\l)$ in eqs.~\eqref{tstbackground2} and~\eqref{tstbackground3} of Appendix~\ref{ap:tstbackground}. The change of variables~\eqref{inverseTuTv3} allows us to express the background fields directly in terms of the deformed energies since the background charges are given by $\Q_{L}(\l) = \frac{c}{6} \cE_L(\l)$ and $\Q_{R}(\l) = \frac{c}{6} \cE_R(\l)$. We then note that eqs.~\eqref{inverseTuTv3} and~\eqref{floweq} allow us to describe the $T\bar{T}$ flow of an undeformed BTZ black hole with charges~\eqref{btzcharges}, 
  \eq{
  \Q_L(0) = \frac{c}{6}\cE_L(0), \qquad \Q_R(0) = \frac{c}{6} \cE_R(0),
  }
to a deformed black string solution whose energies are given by
  \eq{
  \Q_L(\l) &=  -\frac{c}{24\l} \Bigg [ 1 - \frac{12\l}{c} \Q_{\p_\vp}(0)   - \sqrt{1 + \frac{24 \l}{c}  \Q_{\p_t}(0) + \bigg(\frac{12 \l}{c}\bigg)^2 \Q_{\p_\vp}(0)^2} \,\,\Bigg ],  \label{tstleft} \\
    \Q_{R}(\l) & =- \frac{c}{24\l} \Bigg [1 + \frac{12\l}{c} \Q_{\p_\vp}(0) - \sqrt{1 + \frac{24 \l}{c}  \Q_{\p_t}(0) + \bigg(\frac{12\l}{c}\bigg)^2 \Q_{\p_\vp}(0)^2} \,\,\Bigg ], \label{tstright}
  }
where $\Q_{\p_t}(0) \equiv \Q_{L}(0) + \Q_{R}(0)$ and $\Q_{\p_\vp}(0) \equiv \Q_{L}(0)- \Q_{R}(0)$ are the undeformed energy and angular momentum. Using the holographic dictionary~\eqref{dictionary}, we find that eqs.~\eqref{tstleft} and~\eqref{tstright} reproduce the energies of single-trace $T\bar{T}$-deformed CFTs \eqref{ttbarspectrum4}. 

Although the TsT origin of the background is no longer manifest in terms of the new variables $\cE_L$ and $\cE_R$, this parameterization of the black string has other advantages beyond making the $T\bar T$ flow manifest. In particular, note that eq.~\eqref{inverseTuTv3} satisfies the bound $T_u T_v < 1/2\l$ described in Section~\ref{se:tstthermodynamics} for all positive values of $\cE_L$ and $\cE_R$, and hence the apparent pathologies in the charges~\eqref{charges} and the entropy~\eqref{tstentropy} of the black string are no longer present in terms of these variables. In addition, eq.~\eqref{inverseTuTv3} is valid for any choice of $\phi_0$, including values of $\phi_0$ that may depend on $T_u$ and $T_v$. Finally, it is important to note that the change of variables~\eqref{inverseTuTv3} does not affect the results of  previous sections since the matching of the worldsheet spectrum and the thermodynamics of the black string to $T\bar{T}$-deformed CFTs is independent of how we parametrize the $T_u$ and $T_v$ variables.

We conclude by noting that the NS-NS vacuum discussed in Section~\ref{se:NSvacuum} can be phrased in a very natural and simple way using the parameterization~\eqref{inverseTuTv3}. Indeed, the smoothness condition~\eqref{rhoconstraint} for the NS-NS ground state translates into $\cE_L(0) = \cE_R(0)=-{1/4}$ which implies undeformed energies $\Q_{L}(0) = \Q_{R}(0)= -c/24$ that correspond to the energies of the global AdS$_3 \times S^3$ vacuum. Using eq.~\eqref{floweq}, we then find that the deformed energies parameterized by $\cE_L(\l) = \cE_R(\l) = -\frac{1}{4\l} \big (  1  - \sqrt{1 - 2 \l } \,\big)$ match the energies of the $T\bar{T}$-deformed ground state given in \eqref{ttbarvacuum}. Thus, the background~\eqref{NSvacuum} corresponds to the parametrization of the black string that is aligned with the $T\bar{T}$ flow. Note that the derivation of the NS-NS vacuum was done independently of the change of variables~\eqref{inverseTuTv3} by imposing smoothness on the TsT-transformed background. Hence, the results of Section~\ref{se:NSvacuum} provide additional justification for the parametrization of the black string described in this section.


\subsection{TsT black strings with $\l < 0$ \label{se:superluminal}}
Let us briefly comment on the negative sign of the TsT parameter $\l$ which corresponds to the superluminal sign ($\mu < 0$) of the $T\bar{T}$ deformation in the dual field theory. We first note that when $\l < 0$, the TsT black strings~\eqref{tstbackground} are still identified with thermal states in single-trace $T\bar T$-deformed CFTs. This follows from the fact that the gravitational charges~\eqref{charges}, the physical temperatures~\eqref{temperatures}, and the entropy~\eqref{tstentropy} are valid for both positive and negative values of $\l$. Similarly, the holographic dictionary relating the bulk and boundary charges, temperatures, and entropy works in the same way for either sign of the deformation parameter. 

We have seen in Section~\ref{se:ttbarthermodynamics} that for positive signs of the $T\bar{T}$ deformation, potential complex energy states appear at low energies unless $\mu \le R^2/2k$, and that there is an upper bound on the temperatures given by $T_L T_R \le 1/8\pi^2 \mu k$. Using the holographic dictionary $\l = \mu k /\ell^2$, we have shown that these features of $T\bar{T}$-deformed CFTs can be reproduced from TsT black strings in the bulk side of the correspondence. When the deformation parameter is negative, we can also reproduce the critical energy of $T\bar{T}$-deformed CFTs which corresponds to the maximum values of the energy before the spectrum becomes complex. Requiring the total energy~\eqref{ttbarspectrum4} of the singe-trace $T\bar{T}$-deformed theory to be real, we find that the critical energy on the field theory side is given by
  \eq{
R E (\mu) \le  R E^c (\mu) \equiv  {p R^2 \over 2 |\mu|}. \label{ttbarcriticalenergies}
  }
On the gravity side, if we require the TsT background with $\l < 0$ to have a real dilaton, we obtain the following bound on the product of the $T_u$ and $T_v$ parameters (note that this is the same bound obtained in eq.~\eqref{TuTvbound} when $\l > 0$)
  \eq{
  T_u T_v \le \frac{1}{2|\l|}. \label{TuTvbound2}
  }
This bound leads to non-negative physical temperatures (cf.~eq.~\eqref{temperatures} with $\l < 0$) and, using eq.~\eqref{tsttemperatures1} or~\eqref{tsttemperatures2}, to the following upper bound on the gravitational energy of the TsT background
  \eq{
  \Q_{\p_t} \le \Q_{\p_t}^c \equiv \frac{c}{12|\l|}, \label{tstcriticalenergies}
  }
where, following the discussion of Section~\ref{se:tstmatching}, we have set $\phi_0 = 0$. Using the holographic dictionary~\eqref{dictionary} and $c = 6 p k$, we find that the critical energy of the black string~\eqref{tstcriticalenergies} matches the critical energy in $T\bar{T}$-deformed CFTs~\eqref{ttbarcriticalenergies}.

The critical energy~\eqref{ttbarcriticalenergies} can also be understood in a geometrical way. First, note that the scalar curvature of the TsT black string is given, in the Einstein frame, by
  \eq{
  R = \frac{4\l^2 \sqrt{1 - 4\l^2 T_u^2 T_v^2} (r^2 - 4 T_u^2 T_v^2)}{(1 + 2 \l r + 4 \l^2 T_u^2 T_v^2)^{5/2}}.
  }
Thus, when $\l < 0$, the background features a curvature singularity at 
  \eq{
  r_c = \frac{1+4\l^2 T_u^2 T_v^2}{2|\l|}.
  }
From the metric in \eqref{tstbackground} it is clear that closed timelike curves also develop at radius $r = r_c$, in agreement with the observations made in zero temperature backgrounds in~\cite{Giveon:2019fgr,Chakraborty:2019mdf}. This suggests that, in analogy with the double trace deformation~\cite{McGough:2016lol}, a UV cutoff at $r = r_c$ can be used to avoid these pathologies in the space of black string solutions with fixed $\l$.\footnote{The dependence of the cutoff on $T_u$ and $T_v$ is a consequence of our choice of gauge. In particular, we note that a similar feature is observed in the holographic description of the double-trace $T\bar{T}$ deformation~\cite{McGough:2016lol} when one works in the Fefferman-Graham gauge.} We now note that the bound~\eqref{TuTvbound2} implies an upper bound for the horizon that reads
  \eq{
  r_h=2T_uT_v \le \frac{1}{|\l|}. \label{cutoffT}
  }
The critical energy of the TsT black string~\eqref{tstcriticalenergies} was obtained when the bound on the $T_u$ and $T_v$ parameters~\eqref{TuTvbound2} is saturated, which in turn implies that
  \eq{
  r_c = r_h.
  }
Hence, the critical energy~\eqref{tstcriticalenergies} is reached by black strings whose event horizon coincides with the cutoff radius. Note that black strings with $r_h > 1/ |\l|$ still lie within the cutoff since $r_h \le r_c$ for all values of $T_u$ and $T_v$, but these solutions feature complex values of the dilaton.  

We conclude by noting that the NS-NS vacuum is given by eq.~\eqref{NSvacuum} for both positive and negative values of $\l$. This can be seen by requiring smoothness of the $\l < 0$ solution as in Section~\ref{se:NSvacuum} or by following the $T\bar{T}$ flow of $\l < 0$ states as in Section~\ref{se:ttbarflow}. In contrast to the discussion in Section~\ref{se:NSvacuum}, the ground state solution~\eqref{NSvacuum} is real for all $\l < 0$ but develops both curvature singularities and closed timelike curves at $\rho^2 = \rho_c^2 = 1/ 2|\l|$.


\section{Relationship to other backgrounds} \label{se:hhgik}

In this section we consider the Horne-Horowitz black string constructed in~\cite{Horne:1991gn} and generalize it to include angular momentum. We then show that after a change of coordinates this background corresponds to the TsT black string with $\l = 1/2$, the critical value that guarantees a well-behaved spectrum for all states. As a result, the Horne-Horowitz black string with fixed electric charge is dual to a thermal state in a $T\bar{T}$-deformed CFT. Finally, we show that the background considered by Giveon, Itzhaki and Kutasov in~\cite{Giveon:2017nie} also corresponds to the TsT black string with $\l = 1/2$ after a change of coordinates, a gauge transformation, and a choice of $\phi_0$. This background reproduces the thermodynamics of $T\bar{T}$  but with thermodynamic potentials that differ from those derived in Section~\ref{se:tstbtz}.

\subsection{The Horne-Horowitz black string} \label{se:hornehorowitz}

The Horne-Horowitz black string is a three-dimensional, asymptotically zero-curvature background of string theory supported by NS-NS flux that is obtained by gauging an $SL(2,R) \times U(1)$ WZW model. This construction is similar to the one discussed in Section~\ref{se:tsts} except that the gauged $U(1)$ symmetry corresponds to a different subgroup of $SL(2,R)$. The Horne-Horowitz background is characterized by a mass $M$ and a charge $Q$ such that, when $M > |Q|  >0$, the black string features a regular inner horizon and a timelike singularity. These are also features of four-dimensional Reissner-Nordstr\"om black holes, which makes the Horne-Horowitz background an interesting toy model for asymptotically-flat  charged black holes. As we will see, another motivation to study this class of solutions is that they correspond to thermal states in $T\bar{T}$-deformed CFTs.

Let us begin by generalizing the Horne-Horowitz black string constructed in~\cite{Horne:1991gn} by adding angular momentum. The six dimensional metric, $B$-field, and dilaton are given by
  \eq{
   ds_{\textrm{{HH}}}^2 &= \ell^2 \bigg\{ \tfrac{\phantom{d}\scalebox{1}{$d\rho^2$}}{4\Big[ \rho^2  \Big( \tfrac{M^2 + Q^2 - J^2}{M}\Big) \rho + Q^2 \Big]} \! - \! \bigg(\! 1 - \frac{M}{\rho}\bigg) dt^2 \!+\! \bigg(\!1 - \frac{Q^2 - J^2}{M \rho} \bigg) d\vp^2 \!+\! \frac{2J}{\rho} dt d\vp  \!+\!  d\Omega_3^2 \bigg\}, \notag  \\
  B_{\textrm{HH}} &= \frac{\ell^2}{4} \bigg( \cos\t\, d\phi \we d\psi + \frac{2 Q}{\rho} \,du \we dv - 2 du \we dv \bigg),  \label{hhbackground} \\
  e^{2 \Phi_{\textrm{HH}}} &=  \frac{k}{p}\frac{1}{\rho},  \notag
  }
where $M$, $J$, and $Q$ parametrize the mass, angular momentum, and charge. In contrast to the original black string of~\cite{Horne:1991gn}, we compactify the spatial coordinate $\vp$ such that $u = \vp + t /\ell$ and $v = \vp - t /\ell$ satisfy
  \eq{
  (u,v) \sim (u + 2\pi, v + 2\pi).
  }
This identification guarantees that the dual field theory is defined on the cylinder, a property that is crucial in the analysis of $T\bar{T}$-deformed CFTs. Another important difference with the original Horne-Horowitz background is that we have added a constant shift of the $B$-field (the last term in $B_{{\textrm{HH}}}$). Locally, this shift is a gauge transformation that does not affect the analysis carried out in~\cite{Horne:1991gn} but does play a role in the connection to $T\bar{T}$ deformations. In the covariant formulation of charges (Appendix~\ref{ap:charges}), the energy, angular momentum, and charge are respectively given by
  \eq{
  \Q_{\p_t} = \frac{c}{6} (M - Q), \qquad \Q_{\p_\vp} = \frac{c}{6} J, \qquad Q_e = p Q. \label{hhcharges}
  }
Note that the additional gauge transformation in $B $ shifts the definition of the energy by a term that depends on the charge $Q$ but not on the mass or angular momentum. 
For instance, in the original nonrotating background~\cite{Horne:1991gn} the energy is given by $\Q_{\p_t} = \frac{c}{6}M$. In the sector with fixed charge where $\d Q = 0$, the gauge transformation amounts to shifting the zero point of the energy. 

We now show that the rotating Horne-Horowitz background~\eqref{hhbackground} with $Q > 0$ (for the $Q < 0$ case we need to reverse the orientation of the $B$-field) corresponds to a TsT transformation of the BTZ black hole with $\l = 1/2$. This can be seen by a change of coordinates
  \eq{
  \rho = \bigg[ \frac{(M + Q)^2 - J^2}{4M} \bigg] r + \frac{M^2 + Q^2 - J^2}{2 M},\label{radialhh}
  }
whereupon the Horne-Horowitz background~\eqref{hhbackground} reduces to the TsT black string~\eqref{tstbackground} after the following identification of variables\footnote{Alternatively, this can be seen from the parametrization of the TsT black string given in Appendix~\ref{ap:tstbackground} after setting  $\l = 1/2$ and letting $\cE \to M/Q - 1$, $\J \to J/Q$, and $e^{2\phi_0} \to Q$.}
  \eq{
  T_u^2 = \frac{M + J - Q}{M - J + Q}, \qquad T_v^2 = \frac{M - J - Q}{M + J + Q}, \qquad e^{2\phi_0} = Q, \qquad \l = \frac{1}{2}. \label{hhmap}
  }
Let us comment on the space of black string solutions parametrized by~\eqref{hhbackground}. As discussed in Section~\ref{se:tstthermodynamics}, $\l = 1/2$ is the critical value of the deformation parameter for which the spectrum of winding strings, as well as the spectrum of of the dual $T\bar{T}$-deformed CFT, are well-defined for all values of the energies including the ground state. We also note that in the Horne-Horowitz parametrization of the black string the expressions for the conserved charges and the entropy are both regular, and $Q$ is allowed to be positive or negative. 
  The existence of the horizon requires that $(M-|Q|+J) (M-|Q|-J)\le0$, which is equivalent to non-negativity of both the left and the right-moving energies $\Q_L  = \frac{1}{2}(\Q_{\p_t} + \Q_{\p_\vp})$ and $\Q_R  = \frac{1}{2}(\Q_{\p_t} - \Q_{\p_\vp})$.  Relatedly, we see from eq.~\eqref{hhmap} that the Horne-Horowitz black string features an event horizon provided that the background before the TsT transformation corresponds to a BTZ black hole at finite temperature. In analogy with~\cite{Horne:1991gn}, solutions with $M -|Q| < |J|$ do not have a horizon and in general feature conical singularities. One possible exception is the analog of the ground state we discussed in Section~\ref{se:NSvacuum}. However, with the identification $\varphi\sim \varphi+2\pi$, the would-be smooth solution corresponds to $M = J = 0$, which makes the metric~\eqref{hhbackground} ill defined.\footnote{As noted in~\cite{Horne:1991gn}, we can choose an identification that depends on $M$ and $Q$ to make the solution smooth.} In order to describe the NS-NS ground state, we need to go back to the $r$ coordinate using eq.~\eqref{radialhh} followed by another radial redefinition to write the metric in the form~\eqref{NSvacuum} with $e^{2\Phi_{HH}}=0$.  
 
We also note that the parameters appearing in the solution~\eqref{hhbackground} have slightly different origins in the Horne-Horowitz construction and the one presented here. In the former case, the solution is built as a deformation of pure AdS$_3$ featuring no parameters, except the cosmological constant that is being kept fixed, and the constant expectation value $\Phi_0$ of the dilaton. Turning on the deformation introduces another parameter $\kappa$, which can be understood as a combination of $\lambda$ and the temperature. The parameters $(\Phi_0,\kappa)$ are then related to the mass $M$ and charge $Q$ of the black string. In the present paper, the black string is obtained as a deformation of a BTZ black hole, and is characterized by four parameters: the mass and angular momentum of the original solution, the expectation value $\Phi_0$ of the dilaton, and the deformation parameter $\lambda$. The solution~\eqref{hhbackground} is then obtained at fixed $\lambda = 1/2$, which leaves us with three parameters for three conserved charges that depend on $M$, $J$ and $Q$. Further restricting to the non-rotating black strings, we obtain the same two parameter phase space parameterized by $M$ and $Q$. The two approaches are consistent as the deformation parameter in the original Horne-Horowitz background~\cite{Horne:1991gn} can be interpreted as an arbitrary mass parameter at fixed $\lambda$.

Not surprisingly, the thermodynamics of the Horne-Horowitz black string with fixed electric charge $Q$ matches the thermodynamics of single-trace $T\bar{T}$-deformed CFTs with central charge $c_Q \equiv 6 p k Q$. Indeed, the entropy of the Horne-Horowitz background can be written in terms of the left and right-moving energies as
\eq{
S_{_{\textrm{HH}}} & = 2\pi \bigg [ \sqrt{  \Q_{L} \Big ( \frac{c_Q}{6} +  \Q_{R} \Big)} +\sqrt{ \Q_{R} \Big (\frac{c_Q}{6} +  \Q_{L} \Big)}\, \bigg ], \label{hhentropy}
}
which makes the relationship to $T\bar{T}$-deformed CFTs manifest. Thus, we conclude that the Horne-Horowitz black string with fixed electric charge is dual to a thermal state in a single-trace $T\bar{T}$-deformed CFT.


\subsection{The Giveon-Itzhaki-Kutasov black string} \label{se:lstbackground}

Let us now consider the black string studied in refs.~\cite{Giveon:2017nie} (see also~\cite{Hyun:1997jv}) which was argued to reproduce the thermodynamics of $T\bar{T}$-deformed CFTs. We will show that, after a change of coordinates and a gauge transformation, this background corresponds to the non-rotating TsT black string with $\l = 1/2$ and a temperature-dependent value of $\phi_0$. Although $\d Q_e \ne 0$ for this background, the expression for the entropy can be written in terms of the energy in a way that resembles the asymptotic density of states of $T\bar{T}$-deformed CFTs.

The finite-temperature background considered in~\cite{Giveon:2017nie} is given in the string frame by
  \eqsp{
  ds_{\textrm{GIK}}^2 &= \ell^2 \bigg\{ \frac{d\rho^2}{\rho^2 - \rho_0^2} + \frac{4 \rho^2 du dv + \rho_0^2 (du - dv)^2}{4(\rho^2 + \rho_1^2)} + d\Omega_3^2 \bigg\}, \\
  B_{\textrm{GIK}} &= \frac{\ell^2}{4} \bigg( \cos\t\, d\phi \we d\psi + \frac{2 \rho_1 \sqrt{\rho_0^2 + \rho_1^2\,}}{\rho^2 + \rho_1^2} \,du \we dv \bigg),  \\
  e^{2 \Phi_{\textrm{GIK}} } &= \frac{k}{p} \frac{1}{\rho^2 + \rho_1^2}, \label{lstbackground}
  }
where the $u$ and $v$ coordinates satisfy
  \eq{
  (u,v) \sim (u + 2 \pi, v + 2\pi). 
   }
In our conventions the constants $\rho_0^2$ and $\rho_1^2$ are dimensionless with the latter given by $\rho_1^2 = {8 p k G_3}/{\sqrt{k \ell_s^2}} = 2$. Using the covariant formulation of charges, we find that up to a constant the energy is given by
  \eq{
  \Q_{\p_t} = \frac{c}{6} \rho_0^2, \label{lstenergy}
  }
  while the angular momentum vanishes. Crucially, the background features an energy-dependent electric charge,
    \eq{
    Q_e = \frac{c}{6} \rho_1 \sqrt{ \rho_0^2 + \rho_1^2 \,}, \label{lstQe}
    }
which plays an important role in reproducing the thermodynamics of $T\bar{T}$-deformed CFTs.

The GIK black string is closely related to the TsT black string studied in Section~\ref{se:tstbtz}. Indeed, after the change of coordinates
  \eq{
  \rho^2 = \frac{\rho_1^2}{(1 - T^2)^2} \,(r + 2 T^2), \label{tstcc}
  }
we find that the background~\eqref{lstbackground} is given in terms of the TsT metric, $B$-field, and dilaton in eq.~\eqref{tstbackground} by
  \eq{
  ds^2_{\textrm{GIK}} = ds^2 \qquad B_{\textrm{GIK}} = B + \frac{\ell^2}{2} du \we dv, \qquad e^{2\Phi_{\textrm{GIK}}} = \frac{(1 - T^2)^2}{\rho_1^2} e^{2 \Phi},\label{lsttstmap}
  }
where, in addition to $\l = 1/2$, we set $T_u = T_v = T$ and identify\footnote{It is possible to generalize the GIK background to include angular momentum. In this case $\phi_0$ satisfies
  \eqst{
  e^{4 \phi_0} - 2 e^{2\phi_0} \Big( \frac{1 + T_u T_v}{1 - T_u T_v}\Big) - \frac{(T_u - T_v)^2}{(1 + T_u^2)(1 + T_v^2)} = 0,
  }
  and the conserved charges are given by
  \eqst{
  \Q_{L}  = \frac{c}{6}  \frac{(1 + T_v^2) T_u^2}{1 - T_u^2T_v^2} e^{2\phi_0} + \frac{c e^{2\phi_0}}{12} - \frac{c\rho_1^2}{12}, \qquad \Q_{R}  = \frac{c}{6}  \frac{(1 + T_u^2) T_v^2}{1 - T_u^2T_v^2} e^{2\phi_0} + \frac{c e^{2\phi_0}}{12} - \frac{c \rho_1^2}{12}.\
    }
}
  \eq{
  \rho_0^2 = \frac{4 \rho_1^2 T^2}{(1 - T^2)^2}, \qquad e^{2\phi_0} = \rho_1^2 \bigg(\frac{1+ T^2}{1 - T^2}\bigg).
  }

In terms of $T$, the energy of the GIK black string~\eqref{lstenergy} can be written as
  \eq{
  \Q_{\p_t}  = \frac{c}{3} e^{2\phi_0} \frac{(1 + T^2) T^2}{1 - T^4} + \frac{c}{6} e^{2\phi_0} - \frac{c}{6} \rho_1^2. \label{lstenergy2}
    }
This expression does not match the energy $\Q_{L} + \Q_{R}$ of the TsT background computed in Section~\ref{se:tstthermodynamics}, the latter of which corresponds to the first term in eq.~\eqref{lstenergy2}. This follows from the fact that ($i$) there is a shift of the $B$-field between the GIK and TsT backgrounds that is responsible for the second term in eq.~\eqref{lstenergy2}; and ($ii$) there is a constant shift of the energy (the third term) that is not determined when  integrating the infinitesimal charges. The temperature-dependent value of $\phi_0$ is responsible for making the shift of the gauge field contribute to the energy. Otherwise, when $\phi_0$ is constant, as in the Horne-Horowitz black string with fixed electric charge, the shift of the gauge field translates into a constant shift of the energy. The difference between the charges of the GIK and TsT black strings shows that, although related via a gauge transformation and a shift of the dilaton, these backgrounds describe different physical solutions.

The entropy of the GIK background is given in terms of the energy by
  \eq{
  S_{\textrm{GIK}} = \frac{\pi c}{3} \rho_0 \sqrt{\rho_0^2 + \rho_1^2} = 2\pi \sqrt{\Q_{\p_t} \Big(\frac{c}{3} + \Q_{\p_t}    \Big)}, \label{lstentropy}
    }
which matches the entropy of single-trace $T\bar{T}$-deformed CFTs with $R E = \Q_{\p_t}$ and $\lambda = 1/2$, as originally observed in~\cite{Giveon:2017nie}. Note that the background~\eqref{lstbackground} features an electric charge that depends on the energy. Hence, we expect the holographic dictionary relating the bulk and boundary quantities to differ from our discussion in Section~\ref{se:tstmatching}, where we matched the sector of solutions with fixed electric charge ($\d Q_e \propto \d \phi_0 = 0$) to $T\bar T$-deformed CFTs.\footnote{Note that the reason why it is possible to reproduce the thermodynamics of $T\bar{T}$ with a dilaton that differs from the one considered in Section~\ref{se:tstthermodynamics} is that expression for the energy has also changed.} In order to illustrate the differences in the holographic dictionary, we first note that the Hawking temperature $T_H$ and the chemical potential $\nu$ are given by
  \eq{
  \ell T_H = \frac{1}{2\pi} \sqrt{\frac{\rho_0^2}{\rho_0^2 + \rho_1^2}}, \qquad \nu = -\frac{\pi c}{3} \frac{\rho_1}{\rho_0}, \label{THawking}
  }
  and lead to a consistent first law of thermodynamics,
  \eq{
  \d S_{GIK} = \frac{1}{\ell T_H} \d \Q_{\p_t} + \nu \d Q_e. \label{lstfirstlaw}
  }
As a consequence of the additional chemical potential, the Hawking temperature $T_H$ cannot be matched to the temperature of $T\bar{T}$-deformed CFTs given in eqs.~\eqref{ttbartemperatures1} and~\eqref{ttbartemperatures2}. The reason for this is that the GIK background~\eqref{lstbackground} describes a one-parameter space of solutions where the energy $\Q_{\p_t}$ and the electric charge $Q_e$ are not independent variables. Within this one-one-parameter  phase space, the correct potential conjugate to the energy can be determined by expressing $Q_e$ in terms of $\Q_{\p_t}$ such that the first law becomes $\d S_{GIK} = \big(1/\ell T^{\textrm{eff}}_H \big) \d \Q_{\p_t}$ where $T^{\textrm{eff}}_H$ is the effective temperature. The latter is given by
 \eq{
  T^{\textrm{eff}}_H =  \frac{2\rho_0^2 + 2 \rho_1^2}{2\rho_0^2 + \rho_1^2} T_H = \frac{1}{\pi} \frac{\rho_0 \sqrt{\rho_0^2 + \rho_1^2}}{2\rho_0^2 + \rho_1^2}.
  }
and matches the temperature of $T\bar{T}$-deformed CFTs.

To summarize, the GIK black string~\cite{Giveon:2017nie} can be embedded into our general TsT black string solution~\eqref{tstbackground} up to an appropriate choice of $\phi_0$ and a constant gauge transformation. In particular, the results of Section~\ref{se:tstmatching} and ref.~\cite{Giveon:2017nie} map different sectors of the phase space~\eqref{tstbackground} to single-trace $T\bar T$-deformed CFTs.


 
 \bigskip

\section*{Acknowledgments}
We are grateful to Nikolay Bobev, Lorenz Eberhardt, Victor Gorbenko, Juan Maldacena, CarloAlberto Ratti, Eva Silverstein, and Herman Verlinde for helpful discussions. LA thanks the Institute for Advanced Study for hospitality during the completion of this project. The work of LA and WS was supported by the National Thousand-Young-Talents Program of China and NFSC Grant No.~11735001. The work of LA was also supported by the International Postdoc Program at Tsinghua University and NFSC Grant No.~11950410499. SD is supported in part by the ARC grant ``Holography, Gauge Theories and Quantum Gravity Building models of quantum black holes'', by IISN -- Belgium (convention 4.4503.15) and benefited from the support of the Solvay Family. SD is a Research Associate of the Fonds de la Recherche Scientifique F.R.S.--FNRS (Belgium).


\bigskip

\appendix

\section{Conserved charges} \label{ap:charges}

In this appendix we collect the formulae necessary for the computation of conserved charges of six-dimensional type IIB supergravity. In our conventions, the action is given by
  \eq{
  S_{\textrm{IIB}} = \frac{1}{8\pi^3 \ell_s^4} \int d^6x \sqrt{|g|} \bigg( R - \p_{\mu} \Phi \p^{\mu} \Phi - \frac{e^{-2 \Phi}}{12} H_{\mu\nu\a}H^{\mu\nu\a} \bigg ), \label{sugra}
  }
where $H = dB$ and $g_{\mu\nu} = e^{-{\Phi}} G_{\mu\nu}$ denotes the metric in the Einstein frame. Let us consider a background solution to the equations of motion and denote by $\xi$ a Killing vector of this background. In the covariant formulation of conserved charges~\cite{Barnich:2001jy,Barnich:2007bf}, the infinitesimal variation of the gravitational charge associated with $\xi$ is given by
  \eq{
   \d {\cal Q}_{\xi} = \frac{1}{2!\,4!} \frac{1}{4 \pi^3 \ell_s^4} \int_{\p\ss} \ve_{\a\b\g\s\mu\nu}  \,K_{\xi}^{\mu\nu} dx^{\a}\we dx^{\b} \we dx^{\g} \we dx^{\s} , \label{deltacharge}
  }
where $\p \ss$ denotes the boundary of a codimension-1 spacelike surface. The antisymmetric tensor $K_{\xi}^{\mu\nu}$ featured in eq.~\eqref{deltacharge} is defined by
  \eq{
  K_{\xi}^{\mu\nu} = k_{\xi,g}^{\mu\nu} + k_{\xi,B}^{\mu\nu} + k_{\xi,\Phi}^{\mu\nu},
  }
where $k_{\xi,f}^{\mu\nu}$ denotes the contribution of the field $f = \{g, B, \Phi\}$ to the variation of the charge. For the action~\eqref{sugra}, the antisymmetric tensors $k^{\mu\nu}_{\xi,f}$ are given by~\cite{Barnich:2001jy,Compere:2007vx}
  \eq{
  \begin{split}
  k_{\xi,g}^{\mu\nu} &= \frac{1}{2} \Big \{  \xi^{\nu}\nabla^{\mu} \d g^{\a}{}_{\a} - \xi^{\nu} \nabla_{\a} \d g^{\a\mu} + \xi_{\a} \nabla^{\nu} \d g^{\a\mu} + \frac{1}{2} \d g^{\a}{}_{\a} \nabla^{\nu} \xi^{\mu} - \frac{1}{2} \d g^{\nu\a} \nabla_{\a}\xi^{\mu} \\
  & \hspace{1cm} + \frac{1}{2} \d g^{\nu\a} \nabla^{\mu} \xi_{\a} - (\mu \lra \nu) \Big \},
  \end{split} \\
    \begin{split}
    k_{\z,B}^{\mu\nu}  & = \frac{1}{4} e^{-2\Phi} \Big \{  \big (2 H^{\mu\nu \l} \d \Phi  - \d H^{\mu\nu\l} \big ) \xi^{\a} B_{\a \l} - H^{\mu\nu\l} \xi^{\a} \d B_{\a \l} - \xi^{\mu} H^{\nu \a \l} \d B_{\a \l} 	\\
  &\hspace{2cm}  +\Big( 2\d g^{\mu \s} H_{\s}{}^{\nu\l} +  \d g^{\l\s}H^{\mu\nu}{}_{\s} - \frac{1}{2} \d g^{\s}{}_{\s} H^{\mu\nu \l}  \Big) \xi^{\a} B_{\a \l}    \\
  &\hspace{2cm}  - \d B^{\mu \l} g^{\nu \s}  \L_{\xi} B_{\s \l}  - (\mu \lra \nu) \Big \}, 
\end{split} \\
k^{\mu\nu}_{\xi,\Phi} &= \big( \xi^{\nu}\nabla^{\mu} \Phi - \xi^{\mu}\nabla^{\nu} \Phi \big) \d \Phi.
}
where $H = dB$, $\d f^{\a_1 \dots \a_n} = g^{\a_1 \l_1} \dots g^{\a_n \l_n} \d f_{\l_1 \dots \l_n}$ for any tensor field $f_{\a_1 \dots \a_n}$, and we will only consider zero mode variations of the background such that $\d \equiv \d T_u \p_{T_u} + \d T_v \p_{T_v} +\d \l \p_{\l}$. The conserved charges are not sensitive to the choice of spacelike surface but they do depend on the global properties of the spacetime. We will assume that the coordinates are identified as in~\eqref{tstidentification}, such that the variation of the charge may be written as
\eq{
   \d \Q_{\xi} = \frac{\ell^4}{4 \pi^3 \ell_s^4} \int_{\p\ss} d\vp d\psi d\t d\phi \,K_{\xi}^{t r}  , \label{deltacharge2}
  }
where we recall that $t = \frac{1}{2}(u-v)$ and $\vp = \frac{1}{2}(u+v)$. 

The BTZ and TsT black strings also feature electric and magnetic charges which count the number of coincident fundamental strings and NS5 branes generating the original background. The electric and magnetic charges are respectively denoted by $Q_e$ and $Q_m$ and are given by
  \eq{
  Q_e = \frac{1}{4 \pi^2 \ell_s^2} \int_{S^3} e^{-2\Phi} \star H , \qquad Q_m = \frac{1}{4\pi^2 \ell_s^2} \int_{S^3} H. \label{electricchargedef}
  }
  %
  

\section{Alternative parametrizations of the black string} \label{ap:tstbackground}

In this appendix we consider alternative parametrizations of the TsT black string~\eqref{tstbackground}. These backgrounds are obtained by expressing the $T_u$ and $T_v$ parameters in terms of the $\cE_L(\l)$ and $\cE_R(\l)$ variables
    \eq{
  T_u^2 = \frac{\cE_L(\l)}{1 + 2 \l \cE_R(\l)}, \qquad T_v^2 = \frac{\cE_R(\l)}{1 + 2 \l \cE_L(\l)}, \label{inverseTuTv2}
  }
which are related to the background left and right-moving energies via
  \eq{
  \Q_L = \frac{c}{6} e^{2\phi_0} \cE_L(\l), \qquad \Q_R = \frac{c}{6}  e^{2\phi_0} \cE_R(\l).
  }
Using the following rescaling of the radial coordinate
  \eq{
  r = \frac{\rho}{[1 + 2 \l \cE_L(\l)] [1 + 2 \l \cE_R(\l)]},
  }
we find that the expressions for the metric, $B$-field, and dilaton are given by
 \eq{
   ds^2 &=  \ell^2 \Bigg \{ {d\rho^2\over 4 \big [ \rho^2 - 4 \cE_L(0) \cE_R(0) \big ]} + \frac{\rho  du dv  + \cE_L(\l) [1+2\l \cE_L(\l)] du^2 + \cE_R(\l) [1+2\l \cE_R(\l)] dv^2}{1+ 2\l \big[ \rho+\cE_L(0)+\cE_R(0)\big]} + d \Omega_3^2  \Bigg \}, \notag \\
   B &= \frac{\ell^2}{4} \bigg[ \cos\t\, d\phi \we d\psi  -  \frac{2\rho  + 8 \l \cE_L(\l) \cE_R(\l)}{1+ 2\l \big[ \rho+\cE_L(0)+\cE_R(0)\big]}  \,du \we dv \bigg] ,  \label{tstbackground2} \\ 
  e^{2\Phi} &=  \frac{k}{p}  \bigg[ \frac{1+2 \l [ \cE_L(\l) + \cE_R(\l)]}{1+ 2\l \big[ \rho+\cE_L(0)+\cE_R(0)\big]} \bigg] e^{-2\phi_0}, \notag
    }
where we used the fact that the undeformed variables $\cE_L(0)$ and $\cE_R(0)$ satisfy eq.~\eqref{floweq}. Note that the introduction of $\l$ dependence in the $T_u$ and $T_v$ parameters means that the TsT origin of the black string~\eqref{tstbackground} is no longer manifest in eq.~\eqref{tstbackground2}. As discussed in detail in Section~\ref{se:ttbarflow}, this is necessary to describe the $T\bar{T}$ flow from the BTZ to the TsT black string.

An alternative parametrization of the black string consists of using the change of variables~\eqref{inverseTuTv2} together with the following shift and rescaling of the radial coordinate
  \eq{
  r = \bigg(\frac{1 - 4\l^2 T_u^2 T_v^2}{2\l} \bigg) e^{-2 \phi_0} \rho - \frac{1 + 4 \l^2 T_u^2 T_v^2}{2\l}. \label{generaltstcc}
  }
The change of coordinates~\eqref{generaltstcc} is valid only when $\l \ne 0$. As a result, in this gauge it is not obvious how to recover the background before the TsT transformation. The TsT black string obtained after the change of variables and coordinates given in eqs.~\eqref{inverseTuTv2} and~\eqref{generaltstcc} is given by
  \eqsp{
  ds^2 &= \ell^2 \Bigg \{ \frac{d \rho^2}{4 \Big [ \rho^2 - \Big[ \frac{1 + (1 + 2\l \cE)^2 - 4 \l^2 \J}{1 + 2 \l \cE} \Big] e^{2\phi_0} \rho + e^{4\phi_0} \Big] } - \frac{1}{2\l} \bigg[ 1 - \frac{(1 + 2\l \cE)e^{2\phi_0}}{\rho} \bigg] dt^2  \\
  & \hspace{1.1cm}+ \frac{1}{2\l} \bigg[ 1 - \frac{(1 - 4\l^2 \J^2)e^{2\phi_0}}{(1 + 2 \l \cE) \rho}\bigg] d\vp^2 + \frac{2 \J e^{2\phi_0}}{\rho} dt d\vp + d\Omega_3^2  \Bigg\},  \\
 B &= \frac{\ell^2}{4} \bigg[ \cos\t\, d\phi \we d\psi - \frac{1}{\l} \bigg( 1 - \frac{e^{2\phi_0}}{\rho} \bigg) du \we dv \bigg], \label{tstbackground3} \\
 e^{2\Phi} &= \frac{k}{p} \frac{1}{\rho},
  }
where $\cE \equiv \cE_L(\l) + \cE_R(\l)$ and $\J \equiv \cE_L(\l) - \cE_R(\l)$ are proportional to the deformed energy and angular momentum. In particular, note that when $\l = 1/2$ this background reduces to the generalization of the Horne-Horowitz black string described in Section~\ref{se:hornehorowitz}. 



\ifprstyle
	\bibliographystyle{apsrev4-1}
\else
	\bibliographystyle{JHEP}
\fi

\bibliography{jtbar}



\end{document}
